\documentclass{aa}

\usepackage{graphicx}
\usepackage{amssymb}
\usepackage{amsmath}
\usepackage{longtable}
\begin{document}

\title{Organic molecules in protoplanetary disks around T~Tauri and
   Herbig~Ae stars}

\author{Wing-Fai Thi\inst{1,2,3} \and Gerd-Jan van Zadelhoff\inst{1,4} \and 
Ewine F.\ van Dishoeck \inst{1}}
\institute{Leiden Observatory,
P.O. Box 9513, NL 2300 RA, The Netherlands \and
Department of Physics and Astronomy, University College
  London, Gower Street, London WC1E 6BT, U.K. \and  Sterrenkundig Instituut Anton Pannekoek, Kruislaan 403, 1098 SJ Amsterdam, The Netherlands
\and Koninklijk Nederlands Meteorologisch Instituut, P. O. Box 201, 3730 AE De Bilt, The Netherlands}
\offprints{Ewine F. van Dishoeck, \email{ewine@strw.leidenuniv.nl}}
\date{}
\abstract{The results of single-dish observations of low- and high-$J$
  transitions of selected molecules from protoplanetary disks around
  two T~Tauri stars (\object{LkCa~15} and \object{TW~Hya}) and two
  Herbig~Ae stars (\object{HD~163296} and \object{MWC~480}) are
  reported. Simple molecules such as CO, $^{\rm {13}}$CO, HCO$^+$, CN
  and HCN are detected.  Several lines of H$_{\rm 2}$CO are found
  toward the T~Tauri star \object{LkCa~15} but not in other objects.
  No CH$_{\rm 3}$OH has been detected down to abundances of
  $10^{-9}-10^{-8}$ with respect to H$_{\rm 2}$. SO and CS lines have
  been searched for without success. Line ratios indicate that the
  molecular emission arises from dense (10$^{\rm 6}$--10$^{\rm 8}$
  cm$^{\rm {-3}}$) and moderately warm ($T \sim$~20--40~K)
  intermediate height regions of the disk atmosphere between the
  midplane and the upper layer, in accordance with predictions from
  models of the chemistry in disks. The sizes of the disks were
  estimated from model fits to the $^{12}$CO 3--2 line profiles.  The
  abundances of most species are lower than in the envelope around the
  solar-mass protostar \object{IRAS 16293-2422}.  Freeze-out in the
  cold midplane and photodissociation by stellar and interstellar
  ultraviolet photons in the upper layers are likely causes of the
  depletion. CN is strongly detected in all disks, and the CN/HCN
  abundance ratio toward the Herbig Ae stars is even higher than that
  found in galactic photon-dominated regions, testifying to the
  importance of photodissociation by radiation from the central object
  in the upper layers.  DCO$^+$ is detected toward \object{TW~Hya},
  but not in other objects.  The high inferred DCO$^+$/HCO$^+$ ratio
  of $\sim$0.035 is consistent with models of the deuterium
  fractionation in disks which include strong depletion of CO.  The
  inferred ionization fraction in the intermediate height regions as
  deduced from HCO$^+$ is at least $10^{-11}-10^{-10}$, comparable to
  that derived for the midplane from recent H$_2$D$^+$ observations.
  Comparison with the abundances found in cometary comae is made.

}

\titlerunning{Molecules in protoplanetary disks}

\authorrunning{Thi et al.}  

\maketitle

\section{Introduction}\label{moldisks:intro}

The protoplanetary disk phase constitutes a key period in the
evolution of matter between the young protostellar and the mature
planetary system stages. Before their incorporation into comets and
large bodies, the gas and dust could have participated in a complex
chemistry within the disk. Studies of the chemistry in disks are
therefore important to quantify the chemical composition of
protoplanetary material.

\ \\
The chemical composition of the envelopes around young protostars is
now known with increasing detail thanks to the combination of rapid
advances in detectors and antenna technology and improved models
(e.g., van Dishoeck \& Blake \cite{vDB98}; Langer et al.\ 
\cite{Langer00}).  Part of this gas and dust settles around the
pre-main-sequence star in the form of a disk, and after the collapse
and accretion onto the star cease, planets and comets can form by
accumulating gaseous and solid material on timescales of a few million
years (e.g., Lissauer \cite{Lis93}; Beckwith \& Sargent \cite{BS96};
Wuchterl et al.\ \cite{Wuchterl00}).  Surveys from the near-infrared
to the millimeter wavelength range have shown that a large fraction of
1--10 million year old Sun-like pre-main-sequence stars harbors a disk
in Keplerian rotation (e.g., Beckwith et al.\ \cite{BSCG90}).  The
masses of these disks (0.001 - 0.1 M$_{\sun}$) is sufficient to form
a few giant gaseous planets.

Single-dish and interferometric observations of molecular species
other than CO are starting to reveal the chemistry in disks around
classical T~Tauri stars (Dutrey, Guilloteau \& Gu\'elin
\cite{Dutrey97}; Kastner et al.\ \cite{Kastner97}; Simon, Dutrey \&
Guilloteau \cite{Simon00}; Duvert et al.\ \cite{Duvert00}; van
Zadelhoff et al.\ 2001; Aikawa et al.\ \cite{Aikawa2003}; Qi et al.\
\cite{Qi03}; Dartois et al.\ \cite{Dartois03}; Kessler et al.\
\cite{Kessler03}; Wilner et al. \cite{Wilner03}).  The low-$J$
rotational transitions of simple molecules (HCN, CN, HNC, H$_{\rm
2}$CO, HCO$^+$, CS, ...) are detected, but their abundances relative
to H$_{\rm 2}$ are inferred to be orders of magnitude lower than those
observed in dark clouds.  The prevailing explanation of this depletion
involves a combination of freeze-out of the molecules on grain
surfaces in the cold midplane and their photodissociation by
ultraviolet and/or X-rays in the upper atmosphere of disks (see Aikawa
et al.\ \cite{Aikawa99a}, 2002; Bergin et al.\ \cite{Bergin03}).  The
abundances are enhanced in the intermediate height regions, which are
warm enough for the molecules to remain in the gas phase.
Photodesorption induced by ultraviolet radiation (Willacy \& Langer
\cite{WL00}; Westley et al.\ \cite{Westley95}) or X-rays (Najita,
Bergin \& Ullom \cite{Najita01}) can further populate the upper layers
with molecules evaporated from dust grains.
  
We present here the results of a survey of several low- and high-$J$
molecular transitions observed toward two classical T~Tauri stars
(\object{LkCa~15} and \object{TW~Hya}) and two Herbig~Ae stars
(\object{MWC~480} and \object{HD~163296}) using single-dish
telescopes.  In particular, organic molecules such as H$_{\rm 2}$CO,
CH$_{\rm 3}$OH and HCN and deuterated species were searched for.  The
comparison of the two types of objects allows the influence of the
color temperature of the radiation field on the chemistry to be
studied.  There are several advantages in observing high-$J$
transitions over the lower-$J$ ones.  First, detections of CO
$J\!=\!6\!\rightarrow\!5$ and H$_{\rm 2}$ show the presence of a warm
upper surface layer in protoplanetary disks whose temperature is
higher than the freeze-out temperature of most volatile molecules (van
Zadelhoff et al.\ \cite{Zadelhoff01}; Thi et al.\ \cite{Thi01}).
Combined with the high densities in disks, this allows the mid-$J$
levels to be readily populated. Models of flaring disks predict that
the upper layer facing directly the radiation from the central star
can extend out to large radii (Chiang \& Goldreich \cite{CG97};
D'Alessio et al.\ \cite{A99}).  Second, by observing at higher
frequencies with single dish telescopes, the lines suffer less beam
dilution entailed by the small angular size of disks, typically
1--3\arcsec\ in radius, than at lower frequencies. Also, confusion with
any surrounding low-density cloud material is minimized.

The results for the different molecules are compared to those found
for protostellar objects, in particular the solar-mass protostar IRAS
16293-2422, which is considered representative of the initial cloud
from which the Sun and the solar nebula were formed. This so-called
Class 0 object (Andr\'e et al.\ \cite{Andre2000}) is younger than the
protoplanetary disks studied here, only a few $\times 10^4$ yr, and
its chemistry is particularly rich as shown by the number of species
found in surveys in the (sub)millimeter range (e.g., van Dishoeck et
al.\ \cite{vD95}; Ceccarelli et al.\ \cite{Cecca01}; Sch\"oier et al.\ 
\cite{Schoier02}, Cazaux et al.\ \cite{Cazaux03} and references
therein).  The similarities and differences in the chemical
composition between \object{IRAS 16293-2422} and the protoplanetary
disks can be used to constrain the chemical models of disks.

At the other extreme, the results for disks can be compared with those
found for objects in our solar system, in particular comets.  This
will provide more insight into the evolution of matter from the
protoplanetary disk phase to planetary systems.  Unfortunately, the
chemical composition of the large bodies in our solar system has
changed since their formation 4.6 Gyr ago. For example, solar
radiation triggers photochemical reactions in the atmospheres of
planets, and the release of energy from the radioactive decay of
short-lived elements such as $^{\rm 26}$Al causes solids to
melt. Comets, however, could have kept a record of the chemical
composition of the primitive solar nebula because they spent much of
their time in the cold outer region of the Solar System (the Oort
cloud) since their formation (Irvine et al.\
\cite{Irvine00}, Stern \cite{Stern03}).
Comparison of cometary D/H ratio and the CH$_{\rm 3}$OH abundances
with those in disks are particularly interesting.

This paper is organized as follows.  In \S~2, the characteristics of
the observed objects are summarized. In \S~3, the observational
details are provided. The results are given in \S~4 where a simple
local thermodynamical equilibrium (LTE) and statistical equilibrium
analysis is performed. In this section, we also derive several disk
characteristics by fitting the $^{12}$CO 3--2 lines.  In \S~5, the
molecular abundance ratios are discussed. In particular, the CN/HCN
ratio can trace the photochemistry whereas the CO/HCO$^+$ ratio
is a tracer of the fractional ionization.  Finally, a discussion on
the D/H ratio in the disks compared with that found in comets or other
star-forming regions is presented (see also van Dishoeck et al.\ 
\cite{Dishoeck2003}).

\section{Objects}
\label{moldisks:objects}

The sources were selected to have strong CO $J\!=\!3\!\rightarrow\!2$
fluxes and the highest number of molecular lines detected in previous
observations (Qi \cite{Qi01}; Thi et al.\ \cite{Thi01}, van Zadelhoff
et al.\ \cite{Zadelhoff01}).  \object{LkCa~15} is a solar mass T~Tauri
star located in the outer regions of the Taurus cloud.  Its age is
estimated to be $\sim$10 million years, although Simon et al.\ (2000)
argue for an age of only 3--5 million years.  \object{LkCa~15} is
surrounded by a disk whose mass is estimated to be around 0.03
M$_{\sun}$, although a higher mass has been obtained from the fitting
of its spectral energy distribution (SED) (Chiang et al.\
\cite{Chiang01}).  \object{LkCa~15} is one of the strongest millimeter
emitting sources in the sample of T~Tauri stars surveyed by Beckwith
et al.\ (\cite{BSCG90}) along with GG~Tau and DM~Tau.  

\object{TW~Hya} forms part of a young association of stars that has
been discovered only recently and is located at only $\sim$56 ~pc
(Webb et al.\ \cite{Webb99}). \object{TW~Hya} itself is a classical
isolated T~Tauri star with a high X-ray flux and a large lithium abundance.
Its large H$\alpha$ equivalent width is indicative of active disk
accretion at a rate of $\sim$ 10$^{-8}$ M$_{\sun}$ yr$^{-1}$ (Kastner
et al.\ \cite{Kastner02}).  Despite its relatively high age ($\sim$15
Myr), \object{TW~Hya} is surrounded by a disk of mass $\sim$
3$\times$10$^{-2}$ M$_{\sun}$ (Wilner et al.\ \cite{Wilner00}) seen
nearly face-on (Weintraub et al.\ \cite{Weintraub89}; Krist et al.\
\cite{Krist00}; Zuckerman et al.\ \cite{Zuckerman01}).

\object{MWC~480} and \object{HD~163296} were chosen to be
representative of Herbig~Ae stars.  These two objects have the
strongest millimeter continuum emission, with disk masses similar to
those around the two T Tauri stars.  All selected objects show gas in
Keplerian rotation as revealed by CO interferometric observations (Qi
\cite{Qi01}; Mannings \& Sargent \cite{MS97}).
\begin{table*}[!ht]
\centering
\caption{Stellar characteristics$^a$\label{moldisks:tab_stellar_char}}
\begin{tabular}{lllllllllllllll}
\hline
\hline
\multicolumn{1}{c}{Star}&\multicolumn{1}{c}{SpT}&\multicolumn{1}{c}{$\alpha$}&\multicolumn{1}{c}{$\delta$}&\multicolumn{1}{c}{D}&\multicolumn{1}{c}{M$_*$}&\multicolumn{1}{c}{Log($L_*/L_{\sun}$)}
&\multicolumn{1}{c}{Age}\\ 
 & &\multicolumn{1}{c}{(J2000)} &\multicolumn{1}{c}{(J2000)} & \multicolumn{1}{c}{(pc)}&\multicolumn{1}{c}{(M$_{\sun}$)}& & \multicolumn{1}{c}{(Myr)}\\
\hline
{\object{LkCa~15}}   & K7      & 04 39 17.8  & $+$22 21 03 & 140 & 0.8 &$-$0.27 & 11.7\\
{\object{TW~Hya}}    &  K8Ve   & 11 01 51.91 & $-$34 42 17.0 & 56 &1.0 & $-$0.60 & 7$-$15\\
{\object{HD~163296}} & A3Ve    & 17 56 21.26 & $-$21 57 19.5 &122 &2.4 & $+$1.41 &6.0 \\
{\object{MWC 480}}   & A3ep+sh & 04 58 46.27 & $+$29 50 37.0 &131 &2.2 & $+$1.51 & 4.6\\\hline   
\end{tabular}
\begin{flushleft}
  $^a$ See Thi et al.\ (2001) for references. The ages are highly
  uncertain (see also Simon et al.\ \cite{Simon00}).
\end{flushleft}
\end{table*}
\begin{table*}
\centering
\caption{Disk characteristics\label{moldisks:tab_disks_char}}
\begin{tabular}{llllll}
\hline
\hline
\multicolumn{1}{c}{Star}&\multicolumn{1}{c}{Disk Mass$^a$}&\multicolumn{1}{c}{Radius}&\multicolumn{1}{c}{Diameter}&\multicolumn{1}{c}{Inclination}&\multicolumn{1}{c}{Ref.}\\
 & \multicolumn{1}{c}{(10$^{-2}$M$_{\sun}$)} & \multicolumn{1}{c}{(AU)} &\multicolumn{1}{c}{(\arcsec)} & \multicolumn{1}{c}{(\degr)}\\
\hline
{\object{LkCa~15}}   & 3.3  $\pm$ 1.5 & 425     &  6.2  & \ \ \ 57$\pm$10  & 1\\
{\object{TW~Hya}}    & 3.0  $\pm$ 2.0 & 200     &  7.0  & $<$10      & 2, 3, 4,
5, 6\\
{\object{HD~163296}} & 6.5 $\pm$  2.9 & 310     &  5.0  & \ \ \  32$\pm$5   & 7\\
{\object{MWC~480}}   & 2.2 $\pm$  1.0 & 695     & 10.4  & \ \ \  30$\pm$5   & 7\\      
\hline
\end{tabular}
\begin{flushleft}
  $^a$ Total gas + dust mass computed from millimeter continuum flux
  using a dust opacity coefficient
  $\kappa_{\lambda}$=0.01(1.3mm/$\lambda$) cm$^2$ g$^{-1}$ and
  assuming a standard gas-to-dust mass ratio of 100\\
  References.\ (1) Qi et al.\ (\cite{Qi03}); (2) Weintraub et al.\ 
  (\cite{Weintraub89}); (3) Krist et al.\ (\cite{Krist00}); (4)
  Weinberger et al.\ (\cite{Weinberger02}); (5) Wilner et al.\ 
  (\cite{Wilner00}); (6) Calvet et al.\ (\cite{Calvet02}); (7) Mannings et al.\ (\cite{Mannings96})
\end{flushleft}
\end{table*}

The stellar characteristics of the four objects are given in
Table~\ref{moldisks:tab_stellar_char} and the disk characteristics in
Table~\ref{moldisks:tab_disks_char}.  In this paper, inclination is
defined such that $0$\degr \ is a disk viewed pole-on and $90$\degr \ 
edge-on.  Detailed modeling of the SED of these objects suggests that
the disks are in a state of dust settling, especially \object{LkCa~15}
(Chiang et al.\ \cite{Chiang01}).  The objects were chosen to be
isolated from any cloud material to avoid contamination of the
single-dish data.

\section{Observations}
\label{moldisks:observations}

The observations were performed between 1998 and 2000 with the James
Clerk Maxwell Telescope (JCMT)\footnote{\footnotesize The James Clerk
Maxwell Telescope is operated by the Joint Astronomy Centre in Hilo,
Hawaii on behalf of the Particle Physics and Astronomy Research
Council in the United Kingdom, the National Research Council of Canada
and The Netherlands Organization for Scientific Research.}  located on
Mauna Kea for the high-$J$ transitions (850~$\mu$m window) and with
the 30-m telescope of the Institut de Radioastronomie Millim\'etrique
(IRAM) at Pico Veleta for the lower $J$ lines (1 to 3 mm).  At both
telescopes, the observations were acquired in the beam-switching mode
with a throw of 120\arcsec\ at IRAM and 180\arcsec \ at the JCMT in
the azimuth direction. The observations suffer from beam dilution
owing to the small projected sizes of the disks of the order of
5--10\arcsec\ in diameter, compared to the beam size of the JCMT
(14\arcsec\ at 330 GHz) and IRAM (11.3\arcsec\ at 220 GHz).  The data
were reduced and analyzed with the SPECX, CLASS and in-house data
reduction packages.

The JCMT observations made use of the dual polarization B3 receiver
(315--373 GHz) and were obtained mostly in November--December 1999.
The antenna temperatures were converted to main-beam temperatures
using a beam efficiency of $\eta_{\rm mb}$=0.62, which was calibrated
from observations of planets obtained by the staff at the telescope.

The data were obtained in single sideband mode with the image side
band lines reduced in intensity by about 13~dB (i.e. by a factor of
$\sim$ 20). The sidebands were chosen to minimize the system
temperature and to avoid any unwanted emission in the other sideband.
The integration times range from 5 minutes for the bright $^{12}$CO
$J\!=\!3\!\rightarrow\!2$ lines to 8 hours for the faint lines to
reach a r.m.s.\ noise $\delta T_{\rm ther}$ of 10--20 mK after
binning. The backend was the Digital Autocorrelator Spectrometer (DAS)
set at a resolution of $\sim$~0.15-0.27 km s$^{-1}$ (see
Tables~\ref{moldisks:results1} and \ref{moldisks:results2}), and
subsequently Hanning-smoothed to 0.3-0.6 km s$^{-1}$ in spectra where
the signal-to-noise ratio is low.  Pointing accuracy and focus were
regularly checked by observing planets, and was found to be accurate
to better than 3$''$ rms at the JCMT.

The estimated total r.m.s.  error $\delta T$ at the JCMT associated
with each line is given by the relation (e.g., Papadopoulos \&
Seaquist \cite{Papadop98}):
  \begin{equation}
   \left(\frac{\delta T}{T_{\rm mb}}\right)_{\rm tot}=\left[\left(\frac{\delta T}{T_{\rm mb}}\right)_{\rm ther}^2+\left(\frac{\delta T}{T_{\rm mb}}\right)_{\rm drift}^2+\left(\frac{\delta T}{T_{\rm mb}}\right)_{\rm syst}^2\right]^{1/2}
\label{equation_noise}
\end{equation}  
where the first term on the right-hand side of the relation
expresses the ratio between the thermal r.m.s. temperature and the
main-beam peak temperature averaged over $N_{\rm ch}$ channels and
with a baseline derived from $N_{\rm bas}$ channels:
\begin{equation}
  \left(\frac{\delta T}{T_{\rm mb}}\right)_{\rm ther}=\frac{\delta
   T_{\rm ther}}{T_A^*}\left(\frac{N_{\rm bas}+N_{\rm ch}}{N_{\rm
       bas}N_{\rm ch}}\right)^{1/2}
\end{equation}  
where $\delta T_{\rm ther}$ is the thermal noise per channel and
$T_A^*$ is the antenna temperature given in
Tables~\ref{moldisks:results1} and \ref{moldisks:results2}. The
beam-switching method gives extremely flat baselines such that $N_{\rm
  bas} >> N_{\rm ch}$ and
\begin{equation}
\left(\frac{\delta T}{T_{\rm mb}}\right)_{\rm ther}=\frac{\delta T_{\rm ther}}{T_A^*}\frac{1}{\sqrt{N_{\rm ch}}}
\end{equation}  
The data are subsequently binned to $N_{\rm ch}=$ 2 or 4, and a line
detection is claimed whenever
\begin{equation}
\left(\frac{\delta T}{T_{\rm mb}}\right)_{\rm ther} < 0.3
\end{equation}  

The second therm in Eq.~\ref{equation_noise} concerns the drift of the
temperature scale due to all errors of stochastic nature, and
therefore includes any temperature variation of the cold loads or any
fluctuation of atmospheric opacity. Measurements of spectral standard
sources just before or after the source observations allow an estimate
of this drift, which is generally found to be 10--15\% and may be up
to 20--25\% in difficult parts of the atmospheric window (e.g.
H$_2$D$^+$ amd N$_2$H$^+$ lines) depending on the conditions. The
lines for which calibration sources are available from measurements by
the JCMT staff are mentioned in
Table~\ref{moldisks:calibration_sources}\footnote{see also {\tt
    http://www.jach.hawaii.edu/JACpublic/JCMT/
    Heterodyne\_observing/Standards/}}.

The last term encompasses the systematic error, whose main
contributors are the uncertainty in the value of the beam-efficiency
and pointing errors. As noted above, the pointing at the JCMT was
found to be accurate to better than 3\arcsec \ at 345 GHz. Differences
in beam efficiencies and pointing should also be reflected in the
spectral standard observations, which generally agree within 10-15\%
as noted above. This last term is therefore estimated to contribute at
most 10--20\%. Taking into account all possible sources of errors, the
overall calibration uncertainties can be as high as 30--40\% for a
line detected with 3$\sigma$ in a difficult part of the spectrum,
whereas it is of order 20--25\% for high $S/N$ lines for which
spectral standards have been observed.

As discussed by van Zadelhoff et al.\ (\cite{Zadelhoff01}), our
HCO$^+$ $J=4-3$ intensity measured in 1999 is a factor of 3 weaker
than that obtained by Kastner et al.\ (\cite{Kastner97}).  More
recently, we have re-observed the HCO$^+$ line in May 2004 and find
intensities on two days which agree with those of Kastner et al.\ 
within 10--20\%. For comparison, $^{12}$CO 3-2 spectra taken in 1999,
2000 and 2004 are consistent within 10\% with the Kastner et al.
results taken in 1995 with a different receiver, as are the HCN 4-3
and CN 3-2 results.  Thus, only the 1999 HCO$^+$ result appears
anomalously low, perhaps due to unusually large pointing errors during
those observations related to the JCMT $''$tracking error$''$ problem
\footnote{{\tt http://www.jach.hawaii.edu/JCMT/Facility\_description/
    Pointing/problem\_transit.html}}, unless the ion abundance is
variable. We use only the new 2004 data in our analysis. Note that the
H$^{13}$CO$^+$ and DCO$^+$ data were taken only 1 week apart so that
the analysis of the DCO$^+$/HCO$^+$ ratio should not suffer from any
potential long-term variability. Further monitoring of the HCO$^+$
line is warranted.

The IRAM-30m observations were carried out in December 1998 using the
1--3~mm receivers.  The weather conditions were excellent. The three
receivers and a splitable correlator were used to observe
simultaneously lines at 1.3, 2 and 3~mm. The receivers were tuned
single-sideband. Image band rejection was of the order of 10 dB.
Forward ($F_{\rm eff}$) efficiencies were measured at the beginning of
each run and have been found to be consistent with standard values.
We measured $F_{\rm eff}$=0.9, 0.82 and 0.84 at 100, 150 and 230 GHz
respectively.  The derived beam efficiencies ($\eta_{\rm mb}=B_{\rm
  eff}/F_{\rm eff}$) are 0.57, 0.69, and 0.69 at 1, 2 and 3~mm
respectively using {main-beam efficiencies ($B_{\rm eff}$) provided by
  the IRAM staff.  The pointing and focusing accuracy were regularly
  checked to ensure pointing errors $<$~3\arcsec\ (r.m.s.) by
  observing planets and quasars.  \object{TW~Hya} is unfortunately
  located too far south to observe with the IRAM 30m telescope.

\begin{table*}
\centering
\caption{Antenna temperatures and noise per channel width 
for the T~Tauri stars \label{moldisks:results1}}
\begin{tabular}{lllll}
\hline
\hline
\noalign{\smallskip}
            & \multicolumn{1}{c}{Telescope} & \multicolumn{1}{c}{$dv^a$} & \multicolumn{1}{c}{$T_{\rm A}^*$} &  \multicolumn{1}{c}{$\delta T_{\rm ther}^a$}\\
            &             &       \multicolumn{1}{c}{(km s$^{-1}$)} & \multicolumn{1}{c}{(K)} & \multicolumn{1}{c}{(K)}\\
\hline
\noalign{\smallskip}
\multicolumn{5}{c}{\object{LkCa~15}}\\
\hline
\noalign{\smallskip}
$^{12}$CO $J\!=\!2\!\rightarrow\!1$                  & IRAM30m &       0.10  &\phantom{$<$}0.402 & 0.159 \\ 
$^{12}$CO $J\!=\!3\!\rightarrow\!2$                  & JCMT    &       0.13  &\phantom{$<$}0.258  & 0.051 \\  
$^{13}$CO $J\!=\!3\!\rightarrow\!2$                  & JCMT    &       0.28  &\phantom{$<$}0.074  & 0.026 \\      
C$^{18}$O $J\!=\!2\!\rightarrow\!1$                  & JCMT    &       0.21  & $<$0.084 (2$\sigma$) & 0.042 \\    
C$^{18}$O $J\!=\!3\!\rightarrow\!2$                  & JCMT    &       0.14  & $<$0.054 (2$\sigma$)  & 0.027 \\     
HCO$^+$ $J\!=\!4\!\rightarrow\!3$                    & JCMT    &       0.26  &\phantom{$<$}0.055  & 0.018 \\     
H$^{13}$CO$^+$ $J\!=\!4\!\rightarrow\!3$             & JCMT    &      0.13  &  $<$0.078 (2$\sigma$)  & 0.039 \\  
DCO$^+$ $J\!=\!5\!\rightarrow\!4$                    & JCMT    &       0.13  &  $<$0.048 (2$\sigma$)  & 0.024 \\     
CN $J\!=\!3\frac{7}{2}\!\rightarrow\!2\frac{5}{2}$   & JCMT    &       0.27  &\phantom{$<$}0.070  & 0.018 \\        
HCN $J\!=\!4\!\rightarrow\!3$                        & JCMT    &       0.26  &\phantom{$<$}0.051  & 0.022 \\     
CS $J\!=\!7\!\rightarrow\!6$                         & JCMT    &       0.27  &  $<$0.038 (2$\sigma$)  & 0.019 \\     
H$_{\rm 2}$CO $J\!=\!2_{12}\!\rightarrow\!1_{11}$    & IRAM30m &       0.16  &\phantom{$<$}0.021 & 0.008 \\     
H$_{\rm 2}$CO $J\!=\!3_{03}\!\rightarrow\!2_{02}$    & IRAM30m &       0.10  &\phantom{$<$}0.025 & 0.021 \\     
H$_{\rm 2}$CO $J\!=\!3_{22}\!\rightarrow\!2_{21}$    & IRAM30m &       0.10  &   $<$0.036 (2$\sigma$) & 0.018 \\     
H$_{\rm 2}$CO $J\!=\!3_{12}\!\rightarrow\!2_{11}$    & IRAM30m &       0.10  &\phantom{$<$}0.066 & 0.016 \\     
H$_{\rm 2}$CO $J\!=\!5_{15}\!\rightarrow\!4_{14}$    & JCMT    &       0.27  &   \phantom{$<$}0.030   & 0.016 \\     
CH$_{\rm 3}$OH $J\!=\!2_K\!\rightarrow\!1_K$            & IRAM30m &       0.12  &   $<$0.014 (2$\sigma$) & 0.007 \\     
CH$_{\rm 3}$OH $J\!=\!4_{\rm 2}\!\rightarrow\!3_1$ E$^+$& IRAM30m &       0.11  &   $<$0.036 (2$\sigma$) & 0.018 \\     
CH$_{\rm 3}$OH $J\!=\!5_K\!\rightarrow\!4_K$            & IRAM30m &       0.09  &   $<$0.036 (2$\sigma$) & 0.018 \\     
N$_{\rm 2}$H$^+$  $J\!=\!4\!\rightarrow\!3$          & JCMT    &       0.12  &  $<$0.454 (2$\sigma$)  & 0.227 \\     
H$_{\rm 2}$D$^+$  $J\!=\!1_{10}\!\rightarrow\!1_{11}$& JCMT    &       0.12  &  $<$0.372 (2$\sigma$)  & 0.186 \\                                           
\hline
\noalign{\smallskip}
\multicolumn{5}{c}{\object{TW~Hya}}\\
\hline
\noalign{\smallskip}
$^{12}$CO $J\!=\!3\!\rightarrow\!2^b$               & JCMT &        0.13   &\phantom{$<$}1.853 & 0.066 \\     
$^{12}$CO $J\!=\!3\!\rightarrow\!2^c$               & JCMT &        0.13   &\phantom{$<$}1.679 & 0.241 \\     

$^{13}$CO $J\!=\!3\!\rightarrow\!2$               & JCMT &        0.14   &\phantom{$<$}0.185 & 0.046 \\     
HCO$^+$ $J\!=\!4\!\rightarrow\!3^b$               & JCMT &        0.13   &\phantom{$<$}0.656 & 0.230 \\     
HCO$^+$ $J\!=\!4\!\rightarrow\!3^c$               & JCMT &        0.13   &\phantom{$<$}1.197 & 0.087 \\     
H$^{13}$CO$^+$ $J\!=\!4\!\rightarrow\!3$          & JCMT &        0.13   &\phantom{$<$}0.050 & 0.016 \\     
DCO$^+$ $J\!=\!5\!\rightarrow\!4$                 & JCMT &        0.13   &\phantom{$<$}0.104 & 0.023 \\     
CN $J\!=\!3\frac{7}{2}\!\rightarrow\!2\frac{5}{2}$& JCMT &        0.14   &\phantom{$<$}0.597 & 0.075 \\     
HCN $J\!=\!4\!\rightarrow\!3$                     & JCMT &        0.13   &\phantom{$<$}0.369 & 0.080 \\     
H$^{13}$CN $J\!=\!4\!\rightarrow\!3$              & JCMT &        0.13   &  $<$0.056 (2$\sigma$) & 0.028 \\     
HNC $J\!=\!4\!\rightarrow\!3$                     & JCMT &        0.13   &  $<$0.089 (2$\sigma$) & 0.049 \\     
DCN $J\!=\!5\!\rightarrow\!4$                     & JCMT &        0.13   &  $<$0.068 (2$\sigma$) & 0.034 \\     
H$_{\rm 2}$CO $J\!=\!3_{12}\!\rightarrow\!2_{11}$ & JCMT &        0.10   &  $<$0.088 (2$\sigma$) & 0.044 \\     
H$_{\rm 2}$CO $J\!=\!5_{15}\!\rightarrow\!4_{14}$ & JCMT &        0.13   &  $<$0.074 (2$\sigma$) & 0.037 \\     
CH$_{\rm 3}$OH $J\!=\!7_K\!\rightarrow\!6_K$      & JCMT &        0.14   &  $<$0.046 (2$\sigma$) & 0.023 \\     
N$_{\rm 2}$H$^+$  $J\!=\!4\!\rightarrow\!3$       & JCMT &        0.12   &  $<$0.710 (2$\sigma$) & 0.355 \\     
H$_{\rm 2}$D$^+$  $J\!=\!1_{10}\!\rightarrow\!1_{11}$  & JCMT &        0.12   &  $<$0.596 (2$\sigma$) & 0.298 \\     
SO $J\!=\!8_8\!\rightarrow\!7_7$                  & JCMT &  0.13 & $<$0.202 (2$\sigma$) & 0.101\\ 
\hline
\end{tabular}
\begin{flushleft}
$^a$ $dv$ is the spectral resolution of the data in km s$^{-1}$; $T_{\rm A}^*$ is the antenna
temperature (not corrected for beam efficiency) and $\delta T_{\rm
ther}$ is the thermal noise per $dv$ channel. $^b$ Data taken in November 1999. $^c$ Data taken in May 2004. 
\end{flushleft}
\end{table*}

\begin{table*}
\centering
\caption{Same than Table~\ref{moldisks:results1} but for the Herbig~Ae stars.\label{moldisks:results2}}
\begin{tabular}{lllll}
\hline
\hline
\noalign{\smallskip}
            & \multicolumn{1}{c}{Telescope} & \multicolumn{1}{c}{$dv$} & \multicolumn{1}{c}{$T_{\rm A}^*$} &  \multicolumn{1}{c}{$T_{\rm ther}$}\\
            &             &       \multicolumn{1}{c}{(km s$^{-1}$)} & \multicolumn{1}{c}{(K)} & \multicolumn{1}{c}{(K)}\\
\hline
\noalign{\smallskip}
\multicolumn{5}{c}{\object{HD~163296}}\\
\hline
\noalign{\smallskip}
$^{12}$CO $J\!=\!3\!\rightarrow\!2$               & JCMT &        0.23   &\phantom{$<$}0.920  & 0.130 \\    
$^{13}$CO $J\!=\!3\!\rightarrow\!2$               & JCMT &        0.28   &\phantom{$<$}0.225  & 0.052 \\   
HCO$^+$ $J\!=\!4\!\rightarrow\!3$                 & JCMT &        0.26   &\phantom{$<$}0.128  & 0.030 \\    
CN $J\!=\!3\frac{7}{2}\!\rightarrow\!2\frac{5}{2}$& JCMT &        0.27   &\phantom{$<$}0.115  & 0.018 \\   
HCN $J\!=\!4\!\rightarrow\!3$                     & JCMT &        0.26   &  $<$0.074 (2$\sigma$)  & 0.037 \\  
H$_{\rm 2}$CO $J\!=\!5_{15}\!\rightarrow\!4_{14}$ & JCMT &        0.27   &  $<$0.204 (2$\sigma$)  & 0.102 \\   
H$_{\rm 2}$CO $J\!=\!2_{12}\!\rightarrow\!1_{11}$ & IRAM30m &     0.17   &  $<$0.018 (2$\sigma$)  & 0.009  \\ 
H$_{\rm 2}$CO $J\!=\!3_{03}\!\rightarrow\!2_{02}$ & IRAM30m &     0.11   &  $<$0.030 (2$\sigma$)  & 0.015  \\ 
H$_{\rm 2}$CO $J\!=\!3_{12}\!\rightarrow\!2_{11}$ & IRAM30m &     0.10   &  $<$0.018 (2$\sigma$)  & 0.009 \\  
CH$_{\rm 3}$OH $J\!=\!2_K\!\rightarrow\!1_K$      & IRAM30m &     0.12   &  $<$0.014 (2$\sigma$)  & 0.007 \\  
\hline
\noalign{\smallskip}
\multicolumn{5}{c}{\object{MWC~480}}\\
\hline
\noalign{\smallskip}
$^{12}$CO $J\!=\!3\!\rightarrow\!2$                   & JCMT &       0.13  & \phantom{$<$}0.498   & 0.073 \\    
$^{13}$CO $J\!=\!3\!\rightarrow\!2$                   & JCMT &       0.28  &\phantom{$<$}0.102   & 0.016 \\    
C$^{18}$O $J\!=\!3\!\rightarrow\!2$                   & JCMT &       0.14  &  $<$0.062 (2$\sigma$)   & 0.031 \\    
HCO$^+$ $J\!=\!4\!\rightarrow\!3$                     & JCMT &       0.26  &\phantom{$<$}0.052   & 0.017 \\    
DCO$^+$ $J\!=\!5\!\rightarrow\!4$                     & JCMT &       0.13  &  $<$0.106 (2$\sigma$)   & 0.053 \\    
CN $J\!=\!3\frac{7}{2}\!\rightarrow\!2\frac{5}{2}$    & JCMT &       0.27  &\phantom{$<$}0.038   & 0.017 \\    
CS $J\!=\!7\!\rightarrow\!6$                          & JCMT &       0.27  &  $<$0.040 (2$\sigma$)   & 0.020 \\    
HCN $J\!=\!4\!\rightarrow\!3$                         & JCMT &       0.53  &  $<$0.024 (2$\sigma$)  & 0.012 \\     
H$_{\rm 2}$CO $J\!=\!3_{12}\!\rightarrow\!2_{11}$     & JCMT &       0.43  &  $<$0.030 (2$\sigma$)  & 0.015 \\     
H$_{\rm 2}$CO $J\!=\!5_{15}\!\rightarrow\!4_{14}$     & JCMT &       0.27  &  $<$0.032 (2$\sigma$)   & 0.016 \\     
H$_{\rm 2}$D$^+$  $J\!=\!1_{10}\!\rightarrow\!1_{11}$ & JCMT &       0.12  &  $<$0.090 (2$\sigma$)   & 0.045 \\    
N$_{\rm 2}$H$^+$  $J\!=\!4\!\rightarrow\!3$           & JCMT &       0.12  &  $<$0.126 (2$\sigma$)   & 0.063 \\    
H$_{\rm 2}$CO $J\!=\!2_{12}\!\rightarrow\!1_{11}$     & IRAM30m &    0.21  &  $<$0.016 (2$\sigma$)   & 0.008 \\    
H$_{\rm 2}$CO $J\!=\!3_{22}\!\rightarrow\!2_{21}$     & IRAM30m &    1.37  &  $<$0.006 (2$\sigma$)   & 0.003 \\    
CH$_{\rm 3}$OH $J\!=\!4_{\rm 2}\!\rightarrow\!3_1$ E$^+$  & IRAM30m    & 1.37  &  $<$0.006 (2$\sigma$)   & 0.003 \\    
\hline
\end{tabular}
\end{table*}
\normalsize
\begin{table*}
\centering
\caption{Calibration sources used for the JCMT observations\label{moldisks:calibration_sources}}
\begin{tabular}{lllll}
\hline
\hline
\noalign{\smallskip}
\multicolumn{1}{c}{Line}&\multicolumn{1}{c}{LkCa15}&\multicolumn{1}{c}{\object{TW~Hya}}&\multicolumn{1}{c}{\object{HD~163296}}&\multicolumn{1}{c}{MWC 480}\\
\hline
\noalign{\smallskip}
$^{12}$CO $J\!=\!3\!\rightarrow\!2$                   &\object{N2071 IR} &\object{N2071 IR}&\object{IRC+10216}&\object{W3(OH)} \\
                                                      &                  &\object{IRC+10216}                 &\object{IRAS16293-2422}&\\
$^{13}$CO $J\!=\!3\!\rightarrow\!2$                   &\object{W3(OH)} &\object{IRC+10216} &\object{NGC6334I}& \object{W3(OH)}\\
                                                      & & &\object{G34.3}& \\
                                                      & & &\object{IRAS16293-2422}\\
C$^{18}$O $J\!=\!2\!\rightarrow\!1$                   & \object{CRL618}&\multicolumn{1}{c}{...} &\multicolumn{1}{c}{...} &\multicolumn{1}{c}{...} \\
C$^{18}$O $J\!=\!3\!\rightarrow\!2$                   &\object{W3(OH)} &\multicolumn{1}{c}{...} &\multicolumn{1}{c}{...} & \multicolumn{1}{c}{...}\\
                                                      & &                  & & \object{N1333IR4}\\
HCO$^+$ $J\!=\!4\!\rightarrow\!3$                     &\object{W3(OH)} &\object{N2071 IR} &\object{G34.3}& \object{W3(OH)}\\
H$^{13}$CO$^+$ $J\!=\!4\!\rightarrow\!3$              &\object{IRC+10216}&\object{IRC+10216}&\multicolumn{1}{c}{...} &\multicolumn{1}{c}{...} \\ 
                                                      &                  &\object{N2071 IR}& & \\ 
DCO$^+$ $J\!=\!5\!\rightarrow\!4$                     &\object{N1333IR4}&\object{IRAS16293-2422}&\multicolumn{1}{c}{...} &\multicolumn{1}{c}{...}\\
                                                      &\object{IRC+10216}&                      & &\\
CN $J\!=\!3\frac{7}{2}\!\rightarrow\!2\frac{5}{2}$    &\object{CRL618} &\object{CRL618} &\object{CRL618} &\object{CRL618} \\
                                                      &                &                &\object{IRAS16293-2422}&\object{G34.3}\\
HCN $J\!=\!4\!\rightarrow\!3$                         &\object{CRL618} &\object{IRC+10216} &\object{IRAS16293-2422} &\object{W3(OH)}\\
                                                      & &                   &                        &\object{CRL618}\\  
                                                      & &                   &                        &\object{OMC1}\\    
HNC $J\!=\!4\!\rightarrow\!3$                         &\multicolumn{1}{c}{...} &\object{N2071 IR} &\multicolumn{1}{c}{...} &\multicolumn{1}{c}{...} \\
DCN $J\!=\!5\!\rightarrow\!4$                         &\multicolumn{1}{c}{...} &\object{N2071 IR}&\multicolumn{1}{c}{...} &\multicolumn{1}{c}{...} \\  
                                                      & &\object{IRC+10216}& & \\
H$^{13}$CN $J\!=\!4\!\rightarrow\!3$                  & \multicolumn{1}{c}{not available}&\multicolumn{1}{c}{...}&\multicolumn{1}{c}{...} &\multicolumn{1}{c}{...}\\                                                    & &\object{IRAS16293-2422}& & \\
CS $J\!=\!7\!\rightarrow\!6$                          &\multicolumn{1}{c}{not observed}&\multicolumn{1}{c}{...} &\multicolumn{1}{c}{...} &\multicolumn{1}{c}{not observed} \\
SO $J\!=\!8_8\!\rightarrow\!7_7$                      &\multicolumn{1}{c}{...} &\multicolumn{1}{c}{...} & \multicolumn{1}{c}{...}&\multicolumn{1}{c}{...} \\
H$_{\rm 2}$CO $J\!=\!3_{12}\!\rightarrow\!2_{11}$     &\multicolumn{1}{c}{...} &\object{N2071 IR} &\multicolumn{1}{c}{...} &\multicolumn{1}{c}{...} \\
H$_{\rm 2}$CO $J\!=\!5_{15}\!\rightarrow\!4_{14}$     &\object{CRL618} &\object{N2071 IR} &\object{IRAS16293-2422} & \object{CRL618}\\
                                                      & &\object{IRC+10216}& & \\
CH$_{\rm 3}$OH $J\!=\!7_K\!\rightarrow\!6_K$          &\multicolumn{1}{c}{...} &\multicolumn{1}{c}{not observed} &\multicolumn{1}{c}{...} &\multicolumn{1}{c}{...} \\
N$_{\rm 2}$H$^+$  $J\!=\!4\!\rightarrow\!3$           &\multicolumn{1}{c}{not available} &\multicolumn{1}{c}{not available} &\multicolumn{1}{c}{...} &\multicolumn{1}{c}{not available} \\
H$_{\rm 2}$D$^+$  $J\!=\!1_{10}\!\rightarrow\!1_{11}$ &\multicolumn{1}{c}{not available} &\multicolumn{1}{c}{not available} &\multicolumn{1}{c}{...} &\multicolumn{1}{c}{not avaialble} \\
\hline
\end{tabular}
\begin{flushleft}
  {\em Note.--} When a observation is carried out over more than one
  run, different standard sources are used depending on their
  availability. The entry ... indicates that the source was not
  observed in that particular line (cf.\ Table
  \ref{moldisks:intensities}), so no standard was needed.
\end{flushleft}
\end{table*}


\section{Results}
\label{moldisks:results}
\subsection{General characteristics}

\begin{figure*}[!ht]
\begin{center}     
\includegraphics[width=15cm]{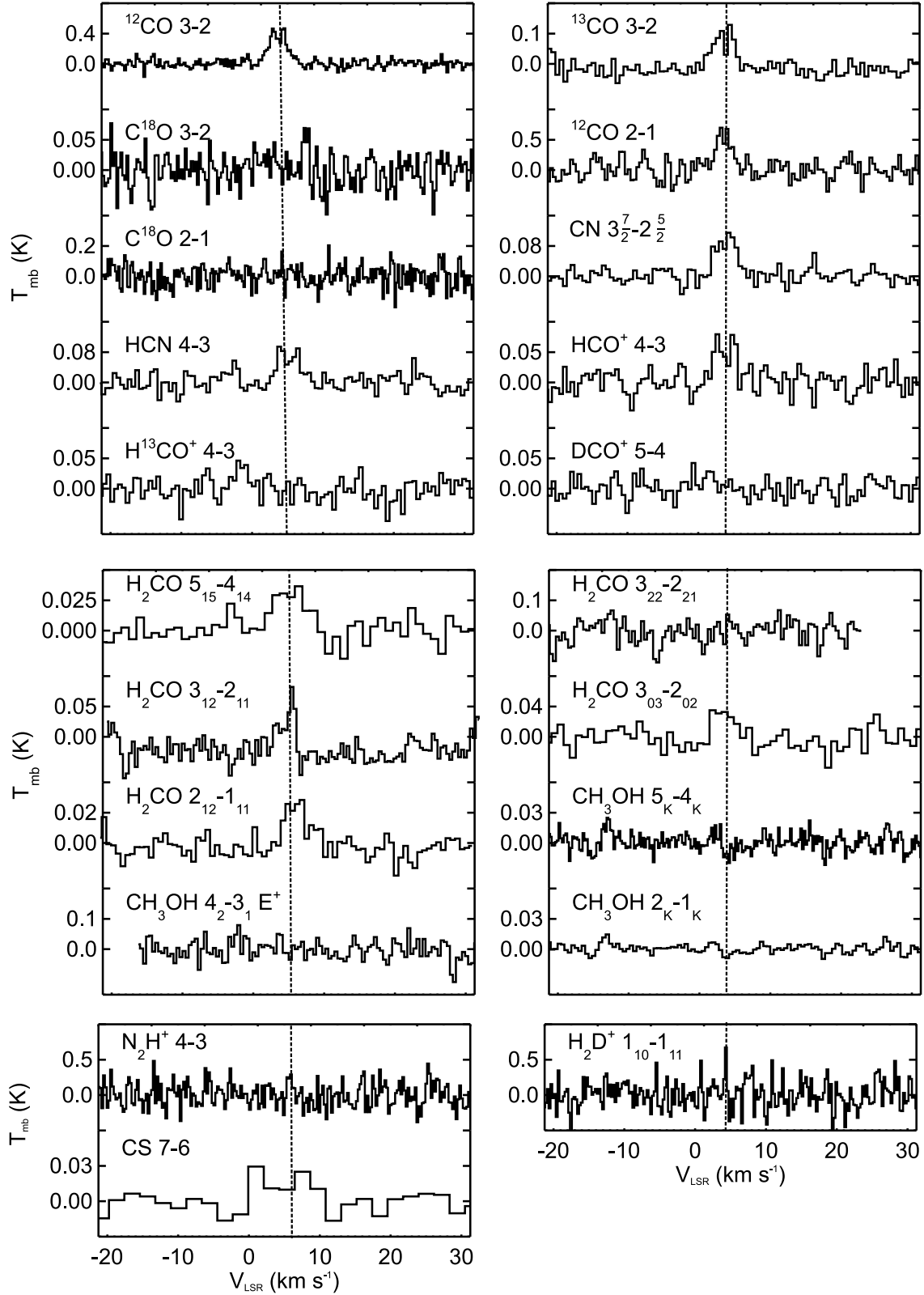}
\end{center}
\caption{Line profiles observed toward LkCa15.  The dashed line
indicates the velocity of the source. Note the different antenna
temperature scales for the different
features.\label{moldisks:fig_lkca152}}
\end{figure*}
The measured antenna temperatures and thermal noise per channel width
are summarized in Tables~\ref{moldisks:results1}
and~\ref{moldisks:results2}.  The spectra are displayed in
Fig.~\ref{moldisks:fig_twhya} to \ref{moldisks:fig_hd163296} on the
main-beam temperature scale for the four sources.  $^{12}$CO
$J\!=\!3\!\rightarrow\!2$ and $^{13}$CO $J\!=\!3\!\rightarrow\!2$ are
detected toward all objects.  Apart from TW~Hya, the profiles of the
$^{12}$CO $J\!=\!3\!\rightarrow\!2$ spectra are double-peaked with
peak separations of $\sim 2$ km s$^{-1}$ for both three objects.  The
$^{12}$CO $J\!=\!3\!\rightarrow\!2$ spectrum of \object{MWC~480} shows
a profile with slightly different peak strengths.  However, the level
of asymmetry is not signifcant compared to the noise.  $^{12}$CO
$J\!=\!3\!\rightarrow\!2$ observations obtained with 30\arcsec offsets
and position-switching to an emission-free position are shown in
Fig.~\ref{moldisks:fig_twhya1} for the four objects.  The maps around
\object{LkCa~15}, \object{TW~Hya} and \object{MWC~480} confirm that
these objects are isolated from cloud material. The observations at
offset positions from \object{HD~163296} show emission at velocities
shifted compared with the velocity of the star.  The extinction to
\object{HD~163296} is sufficiently low that the extended low density
emission is unlikely to arise from a foreground cloud. The offset
emission is only seen in $^{12}$CO, not in $^{13}$CO or other
molecules. Lines arising from high-$J$ transitions require high
critical densities and are therefore not likely to come from a low
density cloud.  A more complete discussion on the possible
contamination by foreground and/or background clouds is given in Thi
et al.\ (\cite{Thi01}).

\begin{figure*}[!ht]
\begin{center}     
\includegraphics[width=15cm]{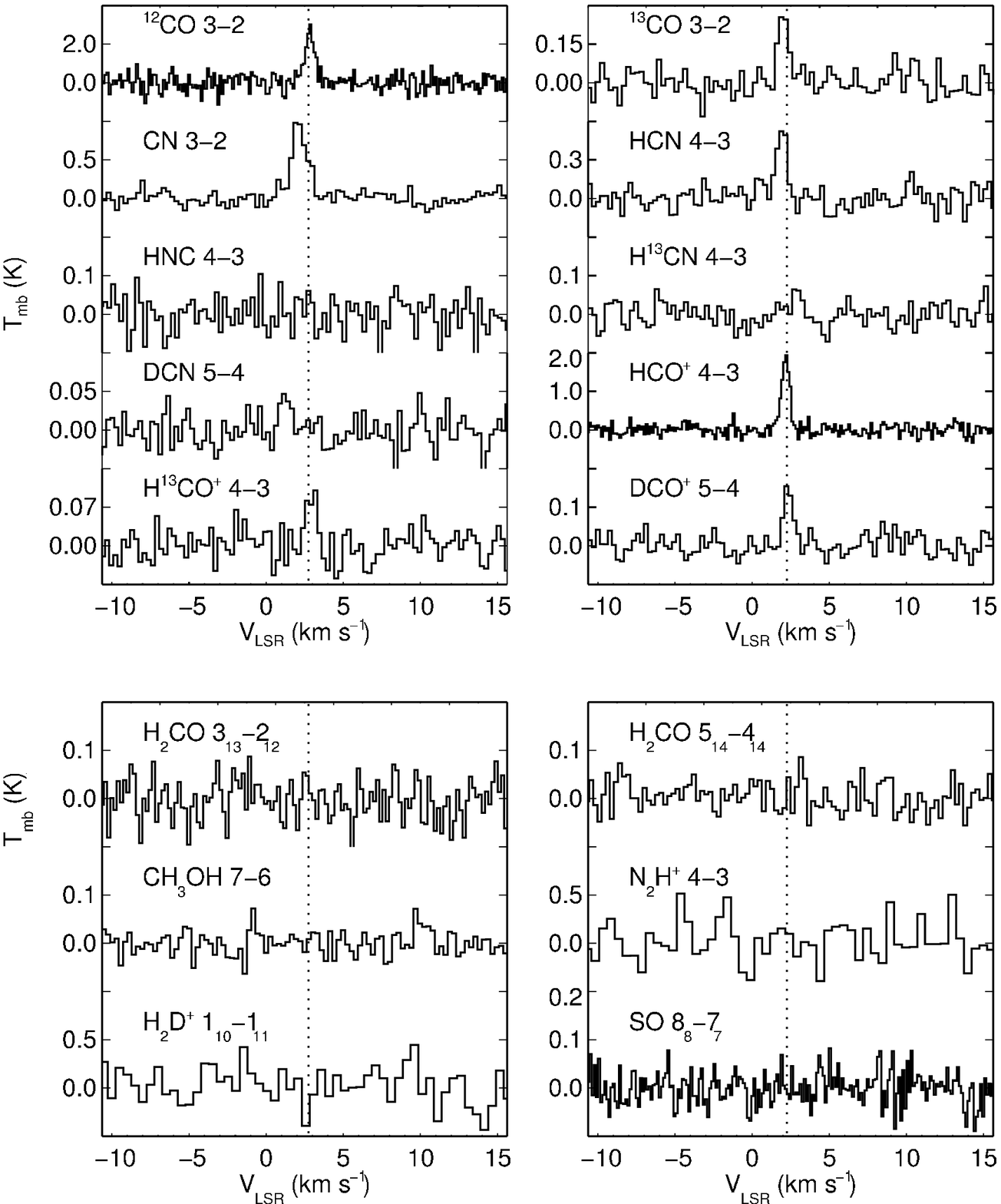}
\end{center}
\caption{Line profiles observed toward \object{TW~Hya}. 
Note the different antenna
temperature scales for the different features.\label{moldisks:fig_twhya}}

\end{figure*}
\begin{figure*}[!ht]
\begin{center}     
\includegraphics[width=15cm]{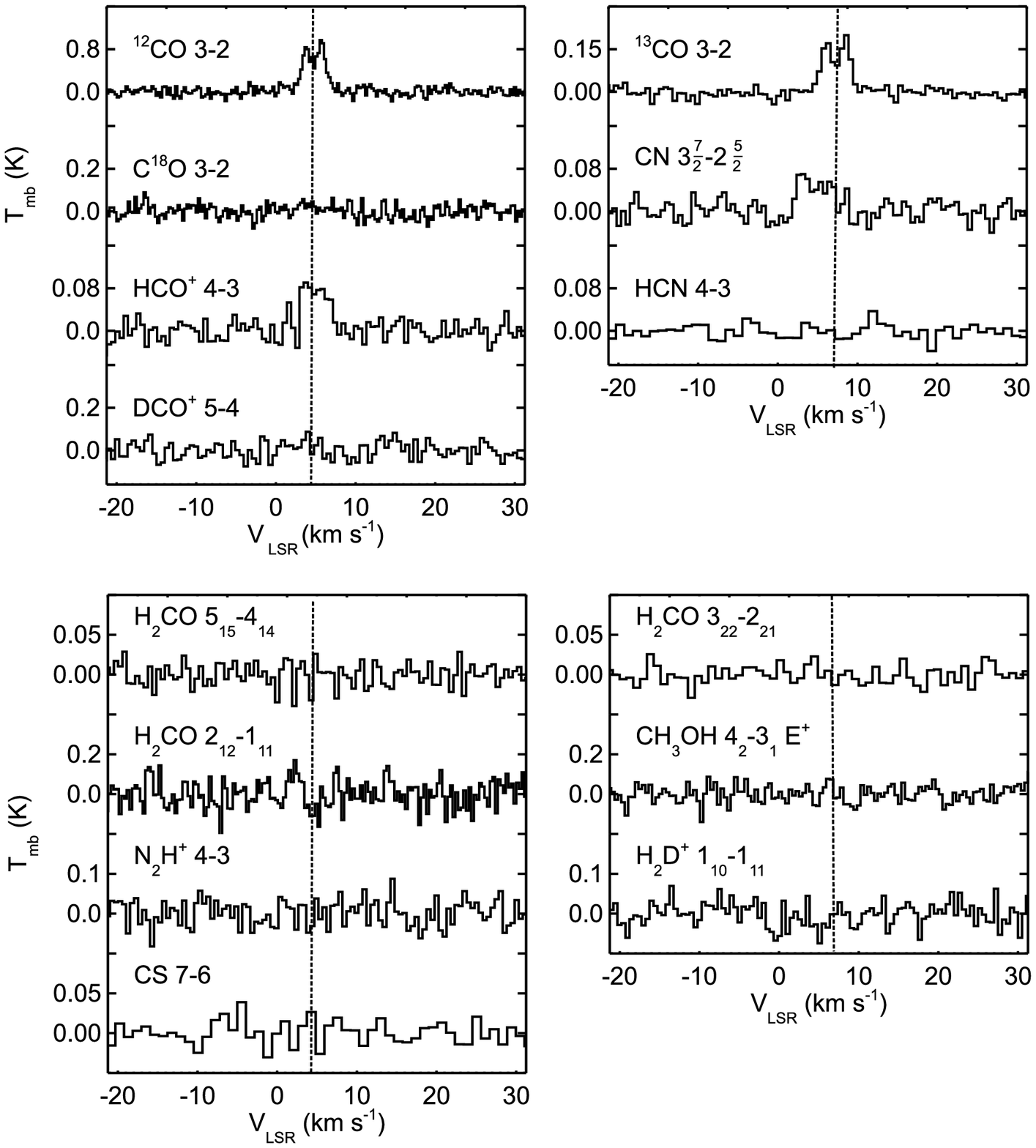}
\end{center}
\caption{Line profiles observed toward MWC 480.
Note the different antenna
temperature scales for the different features.\label{moldisks:fig_mwc480}}
\end{figure*}
\begin{figure*}[!ht]
\begin{center}     
\includegraphics[width=15cm]{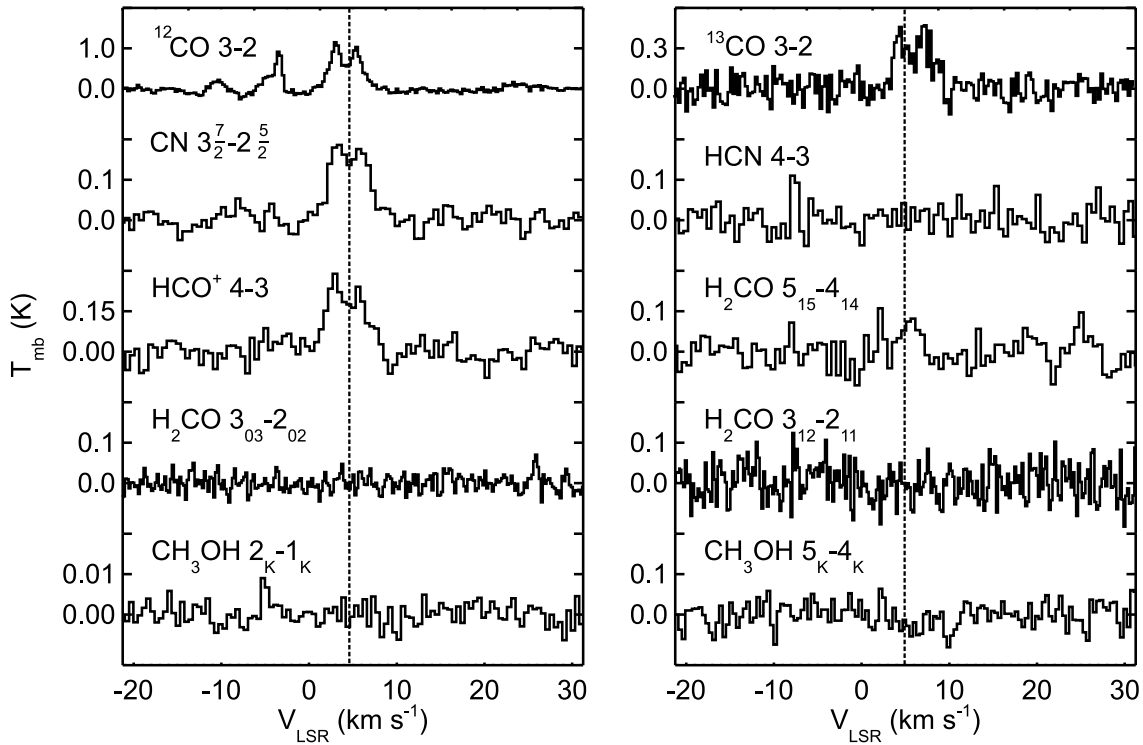}
\end{center}
\caption{Line profiles observed toward \object{HD~163296}.
Note the different antenna
temperature scales for the different features.\label{moldisks:fig_hd163296}}
\end{figure*} 
\begin{figure*}[!ht]
\begin{center}     
\includegraphics[width=15cm]{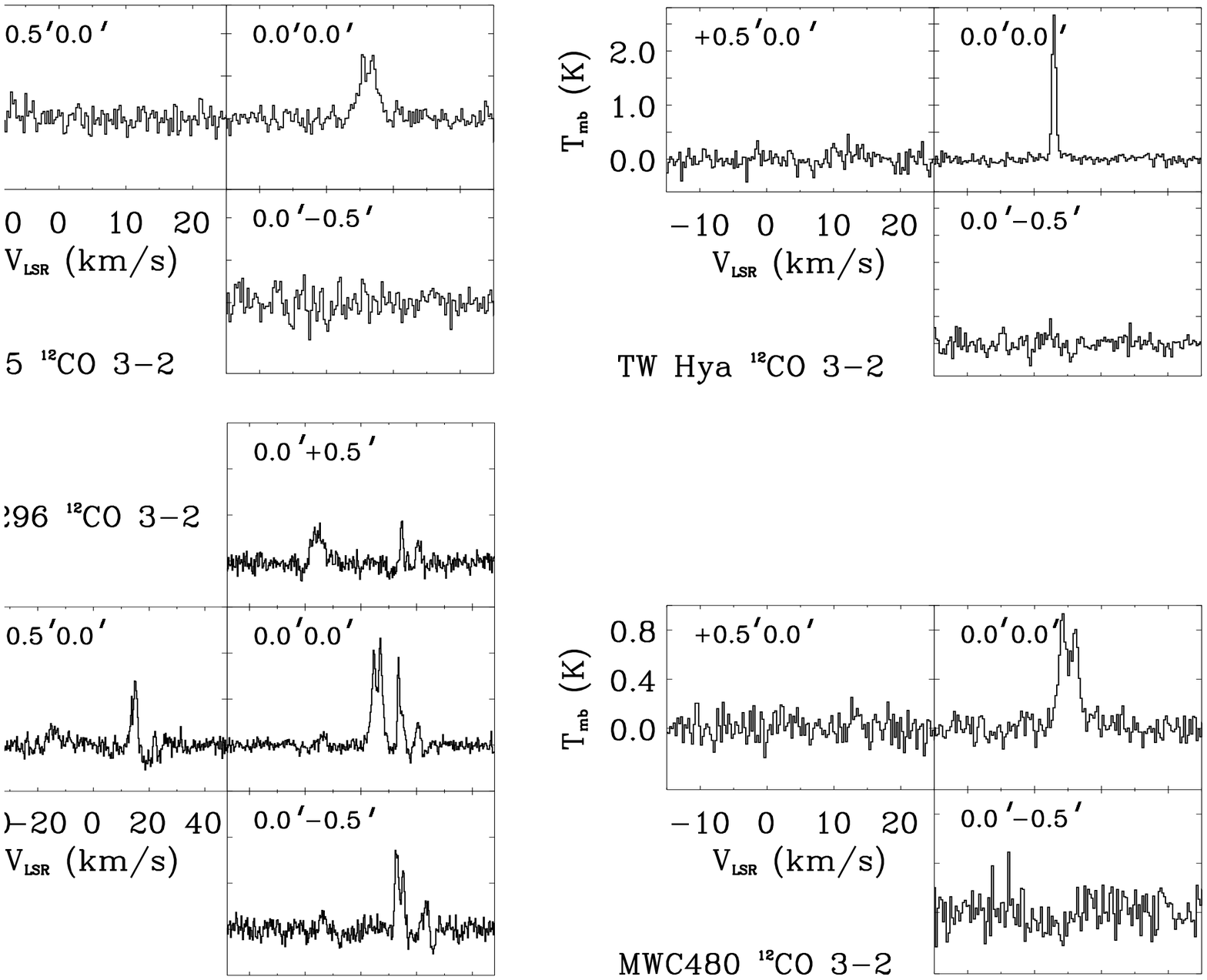}
\end{center}
\caption{$^{12}$CO $J\!=\!3\!\rightarrow\!2$ maps 
  toward \object{TW~Hya}, \object{LkCa~15}, \object{HD~163296} and \object{MWC~480}.\label{moldisks:fig_twhya1}}
\end{figure*}  

High-$J$ lines of various molecules are detected in the disks.  Lines
with high signal-to-noise ratio toward \object{LkCa~15},
\object{HD~163296} and \object{MWC~480} show a double-peak structure
corresponding to emission from a disk in Keplerian rotation viewed
under an inclination angle $i$ (Beckwith \& Sargent \cite{BS93}).  The
line profiles observed toward \object{TW~Hya} are well fitted by a
single gaussian, consistent with a disk seen almost face-on.  The
profiles show no evidence of extended velocity wings characteristic of
molecular outflows in any of the objects.  The velocity integrated
main-beam temperatures for the four sources are summarized in
Table~\ref{moldisks:intensities}. This table includes the energy of
the upper level of the transitions, their critical densities and
frequencies, the telescope at which they were observed and the beam
size. The critical densities are defined as $n_{\rm cr}= A_{\rm ul}/
\sum_{l} q_{\rm ul}$, where $A_{ul}$ is the Einstein A coefficient of
the transition $u\to l$ and $q_{ul}$ the downward rate coefficient.
They have been computed in the optically thin limit at 100~K using the
molecular data listed in Jansen et al.\ (1994) and Jansen (1995).  For
optically thick lines, the critical densities are lowered by roughly
the optical depth of the line.

The upper limits are computed assuming a main beam temperature which
corresponds to twice the r.m.s.\ noise level in a 0.3 km s$^{-1}$ bin
and a line profile similar to that derived from fitting the $^{13}$CO
$J\!=\!3\!\rightarrow\!2$ lines.

The ion HCO$^+$ is detected in all sources.  Toward \object{TW~Hya},
H$^{13}$CO$^+$ is also seen, and the ratio of integrated fluxes
HCO$^+$/H$^{13}$CO$^+$ of 24 is lower than the interstellar isotopic
ratio [$^{12}$C]/[$^{13}$C]~$\sim$~60, indicating that HCO$^+$ line is
optically thick (van Zadelhoff et al.\ \cite{Zadelhoff01}).  Emission
from the CN radical in the $J\!=\!3\!\rightarrow\!2$ line is
surprisingly strong in all objects compared to HCN
$J\!=\!4\!\rightarrow\!3$. HNC $J\!=\!4\!\rightarrow\!3$ was only
searched toward \object{TW~Hya} and has not been detected. The line
intensities for some species in this object differ with previous
observations by Kastner et al.\ (\cite{Kastner97}) (see van Zadelhoff
et al.\ 2001, \S 3). In general, the two Herbig Ae stars display a
less rich chemistry than the two classical T Tauri stars. In
particular, HCN is not detected in either source in our observations.
Qi (\cite{Qi01}), however, reports detection of HCN
$J\!=\!1\!\rightarrow\!0$ toward \object{MWC 480} with the Owens
Valley Millimeter Array (OVRO).

Several lines of H$_{\rm 2}$CO are seen toward \object{LkCa~15} with
the IRAM 30-m and JCMT, but not toward the other three disks. Deep
searches for various CH$_{\rm 3}$OH lines with the IRAM-30m and JCMT
down to very low noise levels did not yield any detections. So far,
CH$_{\rm 3}$OH has only been seen toward \object{LkCa~15} through its
$J\!=\!5_K\!\!\rightarrow\!4_K$ and
$J\!=\!4_{23}\!\!\rightarrow\!3_{13}$ lines using OVRO (Qi
\cite{Qi01}). This illustrates the power of interferometers to detect
minor species in disks, because, with a beam diameter of
1\arcsec--2\arcsec, the fluxes are much less beam-diluted, in addition
to a potentially larger collecting area when a great number of dishes
is available.

No CS $J\!=\!7\!\rightarrow\!6$ line nor lines of SO$_{\rm 2}$, some
of which occur fortuitously in other settings (e.g., near H$_2$CO 351
GHz), were detected toward LkCa~15.  A deep limit on SO was obtained
toward \object{TW~Hya} (Fig.~\ref{moldisks:fig_twhya} and
Table~\ref{moldisks:intensities}).

Finally, DCO$^+$ is detected for the first time in a disk, as reported
by van Dishoeck et al.\ (\cite{Dishoeck2003}). The DCO$^+$
$J\!=\!5\!\rightarrow\!4$ line is observed toward \object{TW~Hya} with
a strength similar to that of H$^{13}$CO$^+$
$J\!=\!4\!\rightarrow\!3$, but the line is not detected toward
\object{LkCa~15} and MWC 480, where H$^{13}$CO$^+$ is also not seen.
There is a hint of a feature near the DCN $J\!=\!5\!\rightarrow\!4$
line toward \object{TW~Hya}, but it is offset by a few km s$^{-1}$.
Deeper integrations or interferometer data are needed to confirm this.
Note also the high critical density needed to excite the DCN
$J\!=\!5\!\rightarrow\!4$ line ($\sim 4.8\ 10^7 $cm$^{-3}$), which may
make it more difficult to detect than DCO$^+$. The ground-state line
of ortho-H$_{\rm 2}$D$^+$ at 372 GHz was searched toward three sources
(\object{LkCa~15}, \object{TW~Hya} and \object{MWC~480}) in a setting
together with N$_2$H$^+$, but neither was detected.  Because of the
poor atmosphere and higher receiver noise at this frequency, the
limits for both H$_2$D$^+$ and N$_2$H$^+$ are not very deep, except
toward \object{MWC 480}. Recently, Ceccarelli et al.\ (2004) have
published the detection of the H$_2$D$^+$ 372 GHz line from the DM Tau
disk using the {\it Caltech Submillimeter Observatory}, together with
a tentative feature from the TW Hya disk. Their integrated line
intensity toward TW Hya is $\int T_{\rm MB} dV=0.39 \pm 0.12$ K km
s$^{-1}$, compared with our 2$\sigma$ limit of 0.20 K km s$^{-1}$.
Taking into account the smaller beam dilution in the JCMT beam and the
measurement uncertainties, the difference between these two data sets
is about a factor of two.

\begin{table*}
\scriptsize
\centering
\caption{Integrated line intensities\label{moldisks:intensities}}
\begin{tabular}{lllllllllll}
\hline
\hline
\noalign{\smallskip}
 &  & & & & & &\multicolumn{4}{c}{$\int \ T_{{\rm mb}}\  {\rm dv}\  ({\rm K\ km \ s}^{-1})$}\\
\noalign{\smallskip}
\cline{8-11}
\noalign{\smallskip}
\multicolumn{1}{c}{Line} &\multicolumn{1}{c}{$E_{\rm upper}$} & \multicolumn{1}{c}{$n_{\rm crit}^a$}& \multicolumn{1}{c}{$\nu$} &\multicolumn{1}{c}{Telescope} & \multicolumn{1}{c}{Beam}& \multicolumn{1}{c}{Cal.$^d$}&\multicolumn{1}{c}{LkCa15}&\multicolumn{1}{c}{\object{TW~Hya}}&\multicolumn{1}{c}{\object{HD~163296}}&\multicolumn{1}{c}{MWC 480}\\
                         &\multicolumn{1}{c}{(K)} &   \multicolumn{1}{c}{(cm$^{-3}$)} & \multicolumn{1}{c}{(GHz)}& & \multicolumn{1}{c}{(\arcsec)}& & & & & \\
\hline
\noalign{\smallskip}
$^{12}$CO $J\!=\!2\!\rightarrow\!1$         & 16.6  & 2.7(3) & 230.538  & IRAM30m & 10.7 &\multicolumn{1}{c}{...} &\phantom{$<$}1.82 & \multicolumn{1}{c}{...} &\multicolumn{1}{c}{...}  &\multicolumn{1}{c}{...} \\  
$^{12}$CO $J\!=\!3\!\rightarrow\!2$         & 33.2  & 8.4(3) & 345.796  & JCMT & 13.7 & yes &\phantom{$<$}1.17 & \phantom{$<$}1.98 & \phantom{$<$}3.78 & \phantom{$<$}2.88\\   
$^{13}$CO $J\!=\!3\!\rightarrow\!2$         & 31.7 & 8.4(3) & 330.587  & JCMT & 14.3 & yes &\phantom{$<$}0.39    & \phantom{$<$}0.24    & \phantom{$<$}0.94 & \phantom{$<$}0.57\\   
C$^{18}$O $J\!=\!2\!\rightarrow\!1$         & 15.8 & 2.7(3) & 219.560  & JCMT & 21.5 & yes &$<$0.20    & \multicolumn{1}{c}{...}&\multicolumn{1}{c}{...} &\multicolumn{1}{c}{...} \\    
C$^{18}$O $J\!=\!3\!\rightarrow\!2$         & 31.6 & 8.4(3) & 329.330  & JCMT & 14.3 & yes &$<$0.14    &  \multicolumn{1}{c}{...}&\multicolumn{1}{c}{...} &\multicolumn{1}{c}{...} \\   
\noalign{\smallskip}
\hline  
\noalign{\smallskip}
HCO$^+$ $J\!=\!4\!\rightarrow\!3$           & 42.8  & 1.8(6) & 356.734  & JCMT & 13.2 & yes &\phantom{$<$}0.26    & \phantom{$<$}1.26& \phantom{$<$}1.10 & \phantom{$<$}0.35\\  
H$^{13}$CO$^+$ $J\!=\!4\!\rightarrow\!3$    & 41.6  & 1.8(6) & 346.998  & JCMT & 13.6 & yes & $<$0.13 & \phantom{$<$}0.07& \multicolumn{1}{c}{...} & \multicolumn{1}{c}{...} \\      
DCO$^+$ $J\!=\!5\!\rightarrow\!4$           & 51.8  & 3.0(6)& 360.169  & JCMT & 13.1 & yes & $<$0.10& \phantom{$<$}0.11 &\multicolumn{1}{c}{...} & \multicolumn{1}{c}{...} \\       
CN $J\!=\!3\frac{7}{2}\!\rightarrow\!2\frac{5}{2}$ & 32.7  & 6.0(6) & 340.248 & JCMT & 13.9 & no & \phantom{$<$}0.67 & \phantom{$<$}1.14 & \phantom{$<$}0.95& \phantom{$<$}0.29\\      
HCN $J\!=\!4\!\rightarrow\!3$               & 42.5 & 8.5(6)& 354.506 & JCMT & 13.3 & yes &\phantom{$<$}0.25 &\phantom{$<$}0.49 & $<$0.20 & $<$0.07\\        
H$^{13}$CN $J\!=\!4\!\rightarrow\!3$        & 41.4 & 8.5(6) & 345.339 & JCMT & 13.6 & no & \multicolumn{1}{c}{...}  &  $<$0.04    & \multicolumn{1}{c}{...} & \multicolumn{1}{c}{...}  \\
HNC $J\!=\!4\!\rightarrow\!3$               & 43.5 & 8.5(6) & 362.630 & JCMT & 13.0 & no & \multicolumn{1}{c}{...}  &$<$0.05 & \multicolumn{1}{c}{...} & \multicolumn{1}{c}{...} \\ 
DCN $J\!=\!5\!\rightarrow\!4$               & 52.1 & 4.8(7)$^b$ & 362.046 & JCMT & 13.0 & yes &\multicolumn{1}{c}{...} &$<$0.03 & \multicolumn{1}{c}{...} & \multicolumn{1}{c}{...} \\ 
CS $J\!=\!7\!\rightarrow\!6$                & 65.8 & 2.9(6) & 342.883 & JCMT & 14.0 & no & $<$0.08 &\multicolumn{1}{c}{...} &\multicolumn{1}{c}{...} & $<$0.08\\          
SO $J\!=\!8_8\!\rightarrow\!7_7$            & 87.7 & 1.8(6) & 344.310 & JCMT & 13.7 & no &\multicolumn{1}{c}{...} &$<$0.10 & \multicolumn{1}{c}{...}& \multicolumn{1}{c}{...}\\
\noalign{\smallskip}
\hline
\noalign{\smallskip}
H$_{\rm 2}$CO $J\!=\!2_{12}\!\rightarrow\!1_{11}$ & 21.9 & 1.0(5) & 140.839    & IRAM30m & 17.5 &\multicolumn{1}{c}{...} &  \phantom{$<$}0.17 & \multicolumn{1}{c}{...}  &$<$0.10  & $<$0.40\\ 
H$_{\rm 2}$CO $J\!=\!3_{03}\!\rightarrow\!2_{02}$ & 21.0 & 4.7(5) & 218.222    & IRAM30m & 11.3 &\multicolumn{1}{c}{...} & \phantom{$<$}0.14 & \multicolumn{1}{c}{...}   &$<$0.30  & \multicolumn{1}{c}{...}\\ 
H$_{\rm 2}$CO $J\!=\!3_{22}\!\rightarrow\!2_{21}$ & 68.1 & 2.3(5) & 218.475    & IRAM30m & 11.3 &\multicolumn{1}{c}{...} & $<$0.10& \multicolumn{1}{c}{...} &\multicolumn{1}{c}{...} & $<$0.06\\ 
H$_{\rm 2}$CO $J\!=\!3_{12}\!\rightarrow\!2_{11}$ & 33.5 & 4.5(5) & 225.697    & IRAM30m & 10.9 &\multicolumn{1}{c}{...} &  \phantom{$<$}0.10 & \multicolumn{1}{c}{...}  &$<$0.30  & \multicolumn{1}{c}{...}\\ 
H$_{\rm 2}$CO $J\!=\!3_{12}\!\rightarrow\!2_{11}$ & 33.5 & 4.5(5) & 225.697    & JCMT & 22.2 & no &  \multicolumn{1}{c}{...} & $<$0.05 & \multicolumn{1}{c}{...}  & \multicolumn{1}{c}{...} \\ 
H$_{\rm 2}$CO $J\!=\!5_{15}\!\rightarrow\!4_{14}$ & 62.5 & 1.7(6) & 351.768    & JCMT & 13.4 & yes &  \phantom{$<$}0.29 &$<$0.04   &$<$0.20  & $<$0.09\\
\noalign{\smallskip}
\hline
\noalign{\smallskip}
CH$_{\rm 3}$OH $J\!=\!2_K\!\rightarrow\!1_K$      & 6.9 & 2.6(3)$^c$ & \phantom{1}96.741     & IRAM-30m & 25.4&\multicolumn{1}{c}{...} &$<$0.05&\multicolumn{1}{c}{...} & $<$0.03 &\multicolumn{1}{c}{...}\\ 
CH$_{\rm 3}$OH $J\!=\!4_{\rm 2}\!\rightarrow\!3_1$ E$^+$& 45.4 & 3.7(4) & 218.440 & IRAM30m & 11.3&\multicolumn{1}{c}{...} &$<$0.10&\multicolumn{1}{c}{...} &\multicolumn{1}{c}{...} & $<$0.20\\
CH$_{\rm 3}$OH $J\!=\!5_K\!\rightarrow\!4_K$      & 34.8 & 4.5(4) & 241.791    & IRAM30m & 10.2&\multicolumn{1}{c}{...} &$<$0.10&\multicolumn{1}{c}{...} & $<$0.10 &\multicolumn{1}{c}{...}\\
CH$_{\rm 3}$OH $J\!=\!7_K\!\rightarrow\!6_K$      & 65.0 & 1.3(5) & 338.409    & JCMT     & 13.9& yes &\multicolumn{1}{c}{...} &$<$0.02& \multicolumn{1}{c}{...} &\multicolumn{1}{c}{...}\\
\noalign{\smallskip}
\hline
\noalign{\smallskip}
N$_{\rm 2}$H$^+$  $J\!=\!4\!\rightarrow\!3$           & 44.7  & 4.4(6) & 372.672   & JCMT &  12.7  & no &$<$0.10&$<$0.30&\multicolumn{1}{c}{...}&$<$0.05\\
H$_{\rm 2}$D$^+$  $J\!=\!1_{10}\!\rightarrow\!1_{11}$ & 104.3 & 1.2(6) & 372.421   & JCMT &  12.7  & no &$<$0.10&$<$0.20&\multicolumn{1}{c}{...}&$<$0.05\\
\noalign{\smallskip}
\hline
\end{tabular}
\normalsize
\begin{flushleft}
  {\em Note.--} The dots indicate {\em not observed}. When a line is
  not detected, a 2$\sigma$ upper limit on $T_{\rm mb}$ in a 0.3 km
  s$^{-1}$ bin is computed and the same profile as the $^{13}$CO
  $J\!=\!3\!\rightarrow\!2$ line is assumed. The beam size ({\em
    HPBW}) is computed for the IRAM-30m using the fitting formula {\em
    HPBW}(\arcsec)=2460/frequency(GHz). $^a$ Unless specified, the
  critical densities are taken from Jansen (\cite{Jansen95b}) for
  $T_{\rm kin}$=100~K assuming optically thin lines. a(b) means a
  $\times$ 10$^{b}$.  $^b$ Computed using the collisional rate
  coefficients for HCN.  $^c$ Derived assuming the collisional rate
  coefficients of Peng \& Whiteoak (\cite{Peng93}). $^d$ Observation
  of calibration sources before and/or after the object. More details
  are given in Table~\ref{moldisks:calibration_sources}.
  \\
\end{flushleft}
\end{table*}

\subsection{Disks properties and molecular abundances}

\subsubsection{Disk mean density}

The mean density can be constrained from line ratios of molecules with high
dipole moments such as HCO$^+$, CN, HCN or H$_{\rm 2}$CO. 
A simple excitation analysis was performed using an escape probability
code described in Jansen et al.\ (\cite{Jansen94}, \cite{Jansen95}).
The code computes the statistical equilibrium population of the
rotational levels given the kinetic temperature, volumn density and
column density. Integrated temperatures of low-$J$ transitions from Qi
(\cite{Qi01}) were used to complement our high-$J$ data. Both sets of
data were corrected for beam dilution by multiplying the observed
velocity integrated main-beam temperatures by the size of the beam as
listed in Table~\ref{moldisks:intensities}.  The analysis for
\object{LkCa~15} and \object{TW~Hya} has been performed previously by
van Zadelhoff et al.\ (\cite{Zadelhoff01}) using HCO$^+$ and HCN, and
takes both the radial and vertical density structure of the disk into
account. Here H$_2$CO is also used as a diagnostic for
\object{LkCa~15} adopting the same method.  Consistent with their
results, we find that the densities in the regions probed by our
observations range from 10$^6$ to 10$^8$ cm$^{-3}$. {\bf This density refers
to the region where the molecular lines are emitted.} The fractions of
mass in a given density interval for various disk models are shown in
Fig.\ 3 of van Zadelhoff et al.\ (\cite{Zadelhoff01}). In all models
(Chiang \& Goldreich \cite{CG97}; D'Alessio et al.\ \cite{A99}; Bell
et al.\ \cite{Bell97}), most of the gas is located in the region of
the disk where the density is greater than 10$^6$ cm$^{-3}$. Those
densities are sufficient for most transitions studied here to be
thermalized. We refer to the paper of van Zadelhoff et al.\ 
(\cite{Zadelhoff01}) for a detailed discussion on the disk models, the
densities derived from line ratios and the disk location where the
lines are expected to become optically thick.

\subsubsection{Disk mean temperature}

The mean kinetic temperatures are less well constrained: the ratio
$^{13}$CO $J\!=\!3\!\rightarrow\!2$ / $^{13}$CO
$J\!=\!1\!\rightarrow\!0$ of 1.35 $\pm$ 0.4 suggests that $T_{\rm
kin}\sim$ 20--40~K for \object{LkCa~15} in the region where the
$^{13}$CO emission originates, assuming that both lines are optically
thin (van Zadelhoff et al.\ \cite{Zadelhoff01}). The bulk material
where CO emits is therefore on average moderately warm and the density
is high enough that the level populations can be assumed to be
thermalized for most cases. The ratios of 2.4 $\pm$ 0.7 for
\object{MWC~480} and 1.7 $\pm$ 0.5 for \object{HD~163296} indicate
similar temperature ranges.  From the CO $J\!=\!6\!\rightarrow\!5$ /
$J\!=\!3\!\rightarrow\!2$ ratios presented in Thi et al.\
(\cite{Thi01}), it is found that the upper layers of disks have
temperatures in the range 25--60~K. The gas temperatures derived from
the H$_{\rm 2}$ data in Thi et al.\ (\cite{Thi01}) are slightly higher
for disks around Herbig~Ae stars than around T~Tauri stars, as
expected if the disks are heated by the radiation from the central
star.  The H$_{\rm 2}$CO
$J\!=\!3_{03}\!\rightarrow\!2_{02}/J\!=\!3_{22}\!\rightarrow\!2_{21}$
ratio is potentially a good temperature indicator, but the
$J\!=\!3_{03}\!\rightarrow\!2_{02}$ line has only been detected toward
\object{LkCa~15}. The limit on the line ratio constrains the
temperature to be below 200~K. A mean temperature of 25~K is adopted
in the remaining parts of the paper.

\subsubsection{Disk size}

The disk sizes are important ingredients for comparing the observed
column densities with models. Since sizes are notoriusly difficult to
derive from low S/N interferometer maps, an attempt has been made to
infer them directly from our model profiles.  Two methods have been
employed. First, since the $^{12}$CO 3--2 emission line is optically
thick, it probes the surface temperature profile of the disk (van
Zadelhoff et al. \cite{Zadelhoff01}). Using the method described by
Dutrey et al.\ (\cite{Dutrey97}) an estimate of the disk size can be
made from the $^{12}$CO $J\!=\!3\!\rightarrow\!2$ lines:

\begin{equation}
\int T_{\rm mb}\ dv=T_{\rm ex}\left(\rho \delta v\right)\left[\frac {\pi \left( R^2_{\rm out} - R^2_{\rm in}\right)} {D^2} \cos i\right]\ \Omega_{\rm a}^{-1}
\label{equa_Dutrey}
\end{equation}

\noindent where $R_{\rm in}$ and $R_{\rm out}$ are the inner and outer
radii, $\delta v$ is the local turbulent velocity (between 0.1 and 0.2
km $^{-1}$) and $\rho$ a geometrical factor of the order of 1.5. We
adopt here $\rho \delta v$ = 0.3 km s$^{-1}$, $R_{\rm in}$=0~AU,
$T_{\rm ex}$=25, 30, and 50~K as the mean disk excitation temperature.
The values for the inclinations $i$, distances $D$ (in AU) and beam
sizes $\Omega_{\rm a}$ are provided in Table
\ref{moldisks:tab_stellar_char} and \ref{moldisks:tab_disks_char}. The
derived disk sizes are given in Table~\ref{tab:fitting_parameters}.
Our estimates for $T_{\rm ex}$=30~K are similar to published values
except for HD~163296, which we find to have twice the size found by
Mannings \& Sargent (\cite{MS97}), who measured it directly from their
$^{12}$CO 1--0 map. Spectra of $^{12}$CO 3--2 emission line were also
generated using a standard parametric disk model as described by,
e.g., Beckwith \& Sargent (\cite{BS93}). The code uses a ray-tracing
method and assumes that the population of the rotational levels is in
Local Thermodynamic Equilibrium. All disks have a power-law density
profile of the form ${n(r)=n_0 (r/1~{\rm AU})^{-2.5}}$. The exact
value of $n_0$ cannot be constrained by fitting optically thick lines
and we assume a typical value of 5 10$^{13}$ cm$^{-3}$ at 1~AU in the
mid-plane. The disk is in hydrostatic equilibrium in the vertical
direction. The inner radius is set to 0.01~AU and is not a significant
parameter.  The free parameters of this model are the excitation
temperature $T_{\rm in}$ {\bf at 1 AU}, the inclination $i$ and the
outer radius $R_{\rm out}$. For simplicity, the gas temperature in the
disk is assumed to have a radial profile power index of 0.5 and
isothermal in the vertical direction.  The best fits are found using a
downhill simplex method (e.g., Press et al.\ \cite{Press97}).
Figure~\ref{moldisks:fig_co_fit} shows the observed spectra and their
best fits obtained with the parameters reported in
Table~\ref{tab:fitting_parameters}. The outer radii found by this
ray-tracing model are smaller than those from the optical depth model
with $T_{\rm ex}=30$~K, which can be ascribed to additional
contributions from warmer gas at large radii not taken account in the
isothermal disk model. Note that $T_{\rm in}$ and $R_{\rm out}$ are
probably degenerate: Table~\ref{tab:fitting_parameters} gives two sets
of values for LkCa~15 that can both fit the spectra. Only high
signal-to-noise spatially resolved interferometer images can lift this
degeneracy.  The larger outer radius (and smaller inner radius
temperature) is adopted, which is closer to that found by direct
fitting of interferometric maps (Qi et al.\ \cite{Qi03}).
The inclinations are consistent with published values (see Table
\ref{moldisks:tab_disks_char}).

\begin{table*}
\small
\begin{center} 
\caption{Disk sizes from the $^{12}$CO $J\!=\!3\!\rightarrow\!2$ 
integrated intensities\label{tab:fitting_parameters}}
\begin{tabular}{llllllll}
\hline
\hline
&\multicolumn{1}{c}{}&\multicolumn{3}{c}{Radius (AU)}&\multicolumn{3}{c}{Disk model}\\
& Literature$^a$& $T_{\rm ex}$=25~K & $T_{\rm ex}$=30~K & $T_{\rm ex}$=50~K & $i$ & $R_{\rm out}^c$ & $T({\rm 1~AU})^b$\\
\hline
\object{LkCa~15}   & 425 & 620 & 559 & 422  & 57 & 290 & 170 \\
\object{LkCa~15}$^d$  & 425 &...&... &... & 57 & 450 & 100 \\
\object{TW~Hya}    & 200 & 238 & 215 & 162  & 3.5& 165 & 140 \\
\object{HD~163296} & 310 & 778 & 701 & 530  & 65 & 680 & 180 \\
\object{MWC~480}   & 695 & 722 & 650 & 492  & 30 & 400 & 170\\
\hline
\end{tabular}
\end{center}
\begin{flushleft}
$^a$ See Table~2 for references \\
$^b$ Gas temperature at 1~AU. The radius of disk is taken to be 0.01 AU. \\
$^c$ Values adopted in subsequent analysis \\
$^d$ Alternative fit to LkCa15 data
\end{flushleft}
\end{table*}
\begin{figure*}[!ht]
\begin{center}     
\includegraphics[width=13cm,angle=90]{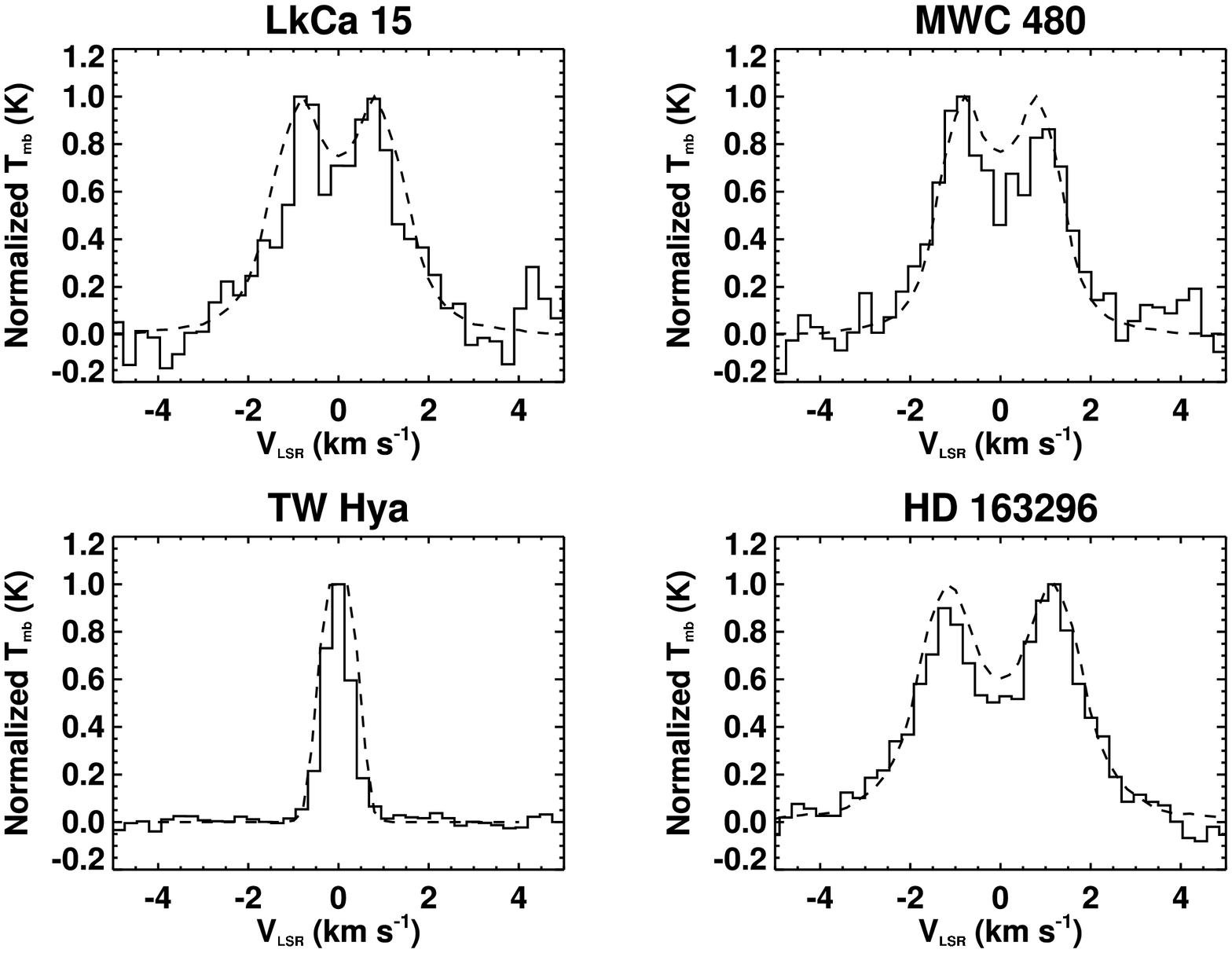}
\end{center}
\caption{Observed (full lines) 
and simulated (dahes lines) $^{12}$CO $J\!=\!3\!\rightarrow\!2$ spectra. 
  The observations are normalized to the peak values.
  The simulations are for a single temperature profile disk model
  (see Table~\ref{tab:fitting_parameters} for parameters).
 \label{moldisks:fig_co_fit}}
\end{figure*}

\subsection{Column densities}

Given the sizes, molecular column densities can be derived from the
observed line strengths.  Two additional assumptions need to be made:
the excitation temperature and the line optical depth.  The line ratio
analysis shows that the lines arise from sufficiently high density
regions (10$^5$--10$^7$ cm$^{-3}$) that they can be assumed to be
thermalized to first order, although small deviations are expected in
the surface layers (see below).  Therefore, a single excitation
temperature of 25~K is adopted to allow easy comparison between the
disks.

The optical depth can be estimated from the ratio of lines from two
isotopologues, assuming that the two species have the same excitation
temperature. Such data are available for a few species and lines, and
the results are summarized in Table~\ref{moldisks:optical_depth}.  It
is seen that both the $^{12}$CO and H$^{12}$CO$^+$ lines are very
optically thick. An alternative method is to compare the size of the
optically thick blackbody which accounts for the line flux to the
actual disk radius derived from the optically thick $^{12}$CO line. We
adopt again the approach of Dutrey et al.~(1997) and rewrite
Eq.(\ref{equa_Dutrey}) as follows:

\begin{equation}
R_{\rm line}({\rm AU})=106.4\left(\frac{\theta_{\rm a}}{1''}\right)\left(\frac{D}{100\ {\rm pc}}\right)\sqrt{\frac{\int T_{\rm mb}\ dv}{T_{\rm ex}\rho \delta v}\times \frac{1}{\pi \cos i}}
\end{equation}
where $\theta_{\rm a}$ is the main-beam diameter at half power in
arcsec. For a Gaussian shape, the solid angle is given by
$\Omega_{\rm a} = 1.133 \theta_{\rm a}^2$.

Assuming $T_{\rm ex}$=25 K, all derived radii are significantly
smaller than the CO disk sizes, except for CN
$J\!=\!3\frac{7}{2}\!\rightarrow\!2\frac{5}{2}$. This would suggest
that the lines from species other than CO are less thick in the outer
parts of the disks ($R>$300~AU) to which our data are most sensitive,
but we consider the direct determination through the isotopologue
ratios more reliable.

\begin{table*}
\small
\centering 
\caption{Optical depths estimates from line ratios between isotopologues}
\label{moldisks:optical_depth}
\begin{tabular}{lllll}
\hline 
\hline
\noalign{\smallskip}
 &  \multicolumn{1}{c}{LkCa15}&\multicolumn{1}{c}{\object{TW~Hya}}&\multicolumn{1}{c}{\object{HD~163296}}&\multicolumn{1}{c}{MWC 480}\\
\noalign{\smallskip}
\hline
$^{12}$CO $J\!=\!3\!\rightarrow\!2$         & \phantom{$<$}26.8  & \phantom{$<$0}8.3   & \phantom{$<$}18.9       & \phantom{$<$}14.9\\
$^{13}$CO $J\!=\!3\!\rightarrow\!2$         & \phantom{$<$0}0.44 & \phantom{$<$0}0.14  & \phantom{$<$0}0.31      & \phantom{$<$0}0.25\\
HCO$^+$ $J\!=\!4\!\rightarrow\!3$           & \phantom{$<$}19.4  & \phantom{$<$0}{\bf 3.8}   & \multicolumn{1}{c}{...} & \multicolumn{1}{c}{...} \\
H$^{13}$CO$^+$ $J\!=\!4\!\rightarrow\!3$    & $<$\ 0.32 & \phantom{$<$0}0.06  & \multicolumn{1}{c}{...} & \multicolumn{1}{c}{...} \\
HCN $J\!=\!4\!\rightarrow\!3$               & \phantom{$<$0}0.75 & $<$\phantom{0}5.1   & \multicolumn{1}{c}{...} & \multicolumn{1}{c}{...}\\
H$^{13}$CN $J\!=\!4\!\rightarrow\!3$        & \phantom{$<$0}0.012& $<$\phantom{0}0.085 & \multicolumn{1}{c}{...} & \multicolumn{1}{c}{...}\\
\noalign{\smallskip}
\hline
\end{tabular}
\begin{flushleft}
  {\em Note.--} We assume that the
isotopologues have the same excitation temperature and that 
[$^{12}$C]/[$^{13}$C]=60.\\
\end{flushleft}
\end{table*}

\normalsize

When the medium is slightly optically thick ($\tau<$3) and $T_{\rm
ex}>>T_{\rm CMB}$ with $T_{\rm CMB}$ = 2.73~K, the column density of
the upper level $N_{\rm u}$ is given by:

\begin{equation}
N_{\rm u}=\frac{8\pi k \nu^2 W}{hc^3A_{ul}}\left(\frac{\Omega_a+\Omega_s}{\Omega_s}\right)\left<\frac{\tau}{1-e^{-\tau}}\right>
\end{equation}
where $\nu$ is the frequency of the transition, $\tau$ is the mean
optical depth, $W=\int T_{\rm mb}\ dv$ is the integrated line
intensity expressed in K km s$^{-1}$, and $\Omega_a$ and $\Omega_s$
are the telescope {\bf main-beam} and the source solid angles,
respectively.  The ratio ($\Omega_a+\Omega_s) / \Omega_s $ is the beam
dilution factor.  
Since the inferred disk sizes are much smaller than the beam sizes,
$(\Omega_a+\Omega_s) / \Omega_s \simeq \Omega_a / \Omega_s $. The
Einstein $A_{\rm ul}$ coefficient of the transition in units of
s$^{-1}$ coefficient is given by:

\begin{equation}
A_{\rm ul}=\left(\frac{64\pi^4\nu^3}{3hc^3}\right)\frac{S\mu^2}{g_{\rm u}}
\end{equation}
where $\mu$ is the dipole moment of the molecule in Debye, $S$ is the
line strength, and $g_{\rm u}= g_{J} g_{K} g_{I}$ is the statistical
weight of the upper level. Finally, the factor
\begin{equation}
\beta^{-1}=\left<\frac{\tau}{1-e^{-\tau}}\right>
\end{equation}
is the escape probability in the so-called Sobolev or
Large Velocity Gradient approximation.    

The column density in level $u$ is related to the total column density $N$
by:
\begin{equation}
N_{\rm u}= \frac{N}{Q_{\rm rot}(T_{\rm ex})} g_{J} g_{K} g_{I}e^{-E_{\rm u}/T_{\rm ex}}
\end{equation}
where $Q_{\rm rot}(T_{\rm ex})$ is the rotational partition function,
$g_{J}$ is the rotational statistical weight factor equal to
$2J + 1$ for diatomic or linear molecules, $g_{K}$ is the
$K$-level degeneracy, $g_{\rm I}$ is the reduced nuclear spin
degeneracy, and $E_{\rm u}$ is the energy of the upper level expressed in
K. For linear molecules, $g_{K}$=$g_{I}$=1 for all
levels. Formaldehyde, H$_{\rm 2}$CO, is an asymmetric top molecule with
$I$=1/2.  Combining the above equations, the total column density can
be expressed as:
\begin{equation}
\begin{split}  
N=\frac{1.67\times10^{14}}{\nu \mu^2 S}Q_{\rm rot}(T_{\rm
  ex})e^{E_{\rm u}/T_{\rm
    ex}}\left<\frac{\tau}{1-e^{-\tau}}\right>\\
\left(\frac{\Omega_a}{\Omega_s}\right)\int T_{\rm mb}\ dv\\
\label{moldisks:equaN}
\end{split}
\end{equation}
This formula is similar to that of Blake et al.\ (\cite{Blake87}) and
Turner (\cite{Turner91}) but with the introduction of the escape
probability and beam dilution factor.

For linear molecules the line strength is equal to the rotational
quantum number $J$. The rotational energy level structure of the two
linear molecules CN and HCN are more complicated than those for CO.
The spin of the unpaired electron for CN (S=$1/2$) and the nuclear
spin of $^{14}$N (I=1) lead to fine- and hyperfine splitting of the
rotational levels. The observed CN 340 GHz line is a blend of three
lines that account for $\simeq$~55\% of all the hyperfine lines
arising from the level $J$=3 (Simon \cite{Simon97}). The line strength
is taken from Avery et al.\ (\cite{Avery92}). An advantage of the
hyperfine splitting is that it decreases the optical depth in each
individual component. The HCN $J\!=\!4\!\rightarrow\!3$ line is also a
blend of hyperfine lines but we assume that all the flux is included
in the observed line. Other constants used to derive the column
densities are taken from existing catalogs (Pickett et al.\ 
\cite{Pickett98}) and are summarized in Thi (\cite{Thi02a}). The
rotational partition functions were calculated using the formulae for
each molecule in Gordy \& Cook (\cite{Gordy84}).   {

The radical CN and the molecule HCN have different critical densities
and the HCN $J\!=\!4\!\rightarrow\!3$ line may be subthermally excited
in the upper layer, so that the inferred $N$(CN)/$N$(HCN) ratio varies
strongly with density. This effect, which can be up to a factor of 2
in the CN/HCN abundance ratio, has been corrected using the
statistical equilibrium calculations described above for the inferred
range of temperatures and densities.  It should be noted that this
correction assumes that the CN and HCN lines come from the same
location inside the disks, which is probably not the case. In disk
models, CN peaks more toward the lower density surface layers than HCN
because CN is mostly formed by radical reactions and photodissociation
of HCN (Aikawa et al.\ \cite{Aikawa2002}).  This effect would lead to
higher CN/HCN abundance ratios than presented here.

Table~\ref{moldisks:column_density} summarizes the beam-averaged
column densities and upper-limits for the observed molecules, adopting
the disk sizes derived from the fits to the $^{12}$CO
$J\!=\!3\!\rightarrow\!2$ spectra using the isothermal disk model (see
parameters in rightmost columns of
Table~\ref{tab:fitting_parameters}).  A single excitation temperature
$T_{\rm ex}=T_{\rm kin}=$ 25~K and an optical depth of $\tau = 1$ are
assumed for all lines.  For optically thick lines with $\tau \gg 1$,
such as those of HCO$^+$ and perhaps HCN, the derived column densities
by these formulae are clearly lower limits. Wherever available, column
densities derived from isotopic data have been used in these cases.

\subsection{Molecular abundances}

It is well known that disk masses derived from CO measurements are
much lower than those obtained from millimeter continuum observations
assuming a gas/dust ratio of 100. Because CO is subject to
photodissociation and freeze-out, one cannot adopt the canonical
CO abundance of CO/H$_2$=$10^{-4}$ found for molecular
clouds; instead, the disk
masses $M_{\rm disk}$ are assumed to be given by the millimeter
continuum observations (see Thi et al.\ 2001,
Table~\ref{moldisks:tab_disks_char}).

It should noted that, in the optically thin limit, the abundances are
independent of the disk size. The derived abundances are summarized in
Table~\ref{tab:mol_abundances}.  As noted above, the abundances
derived from the highly optically thick HCO$^+$ and HCN lines are
likely to be underestimated by up to an order of magnitude.  For
molecules that are detected in all four disks (CN, HCO$^+$), the
abundances vary significantly from object to object.  The
non-detection of HCN toward the Herbig Ae stars confirms the low
abundances in these cases, although the high critical density of the
HCN $J=4\to 3$ line may also play a role.  The upper limits are much
lower in the case of TW~Hya owing to the small distance of this object
and its narrow lines.

The derived abundances and depletion factors are roughly consistent
with the large range of values given by van Zadelhoff et al.\ (2001),
especially for their colder models. A full comparison between the two
studies is difficult since van Zadelhoff et al.\ performed more
detailed radiative transfer modeling with a varying temperature in the
vertical direction. Also, they adopted a smaller disk radius of 200~AU
compared with our value of 450 AU for LkCa~15.  This leads to higher
abundances to reproduce the same line flux, at least for optically
thick lines.

For comparison, we have also re-derived the abundances in the disk
around DM~Tau, where many of the same species have been detected by
Dutrey et al.\ (\cite{Dutrey97}). Their tabulated velocity integrated
flux densities have been converted to velocity integrated main beam
temperatures via the relation:
\begin{equation}
\int T_{\rm mb}({\rm K})\ dv = 10^{-23} \times \int F({\rm Jy}) \ dv
\frac{\lambda^2({\rm cm})}{2k}\Omega_a^{-1}
\end{equation}
where $F$ is the flux density in Jansky, $k$ is the Boltzmann constant
in erg K$^{-1}$, $\lambda$ the wavelength in cm, $\Omega_a$ is the
main-beam solid angle and $dv$ is in km s$^{-1}$.  A total disk mass
of 0.018 M$_{\sun}$ has been used, computed using Equation (6) of Thi
et al.\ (\cite{Thi01}) and a continuum flux of 110~mJy at 1.3~mm
(Guilloteau \& Dutrey \cite{Guilloteau98}).  The abundances are
reported in Table~\ref{tab:mol_abundances}. Our new abundances
estimates are within a factor of 4 of those deduced by Dutrey et al.\ 
(\cite{Dutrey97}) who used a different method to derive their
abundances.

The last column of Table~\ref{tab:mol_abundances} contains the
abundances found in the cold outer region of the protostellar envelope
of IRAS 16293--2422. The latter abundances seem to be higher than
those in disks by at least an order of magnitude, even taking into
account the fact that the disk abundances may be underestimated
because of optical depth effects.   A
noticeable exception is CN, which has a higher abundance in all four
disks and in DM~Tau than in \object{IRAS 16293-2422}.


 \begin{table*}
 \small
 \begin{center}
 \caption{Beam averaged column density of various molecules using
the disk sizes listed in Table~\ref{tab:fitting_parameters}. 
   \label{moldisks:column_density}}
 \begin{tabular}{llllll}
 \hline
\hline  
  &  & \multicolumn{4}{c}{$N  ({\rm cm}^{-2})$}\\
 \noalign{\smallskip}
 \cline{3-6}
 \noalign{\smallskip}
 & &\multicolumn{1}{c}{LkCa15}&\multicolumn{1}{c}{\object{TW~Hya}}&\multicolumn{1}{c}{\object{HD~163296}}&\multicolumn{1}{c}{MWC 480}\\
 \hline
 \noalign{\smallskip}
 \multicolumn{1}{c}{Species} & \multicolumn{1}{c}{Transition} &\multicolumn{1}{c}{$R_{\rm disk}$=450~AU}&\multicolumn{1}{c}{$R_{\rm disk}$=165~AU}&\multicolumn{1}{c}{$R_{\rm disk}$=680~AU}&\multicolumn{1}{c}{$R_{\rm disk}$=400~AU}\\
 \hline
 $^{12}$CO$^{a}$  & $^{13}$CO $J\!=\!3\!\rightarrow\!2$            & \phantom{$<$}1.9(16)  & \phantom{$<$}3.2(16) & \phantom{$<$}3.5(16)& \phantom{$<$}6.9(16)\\
 $^{13}$CO        & $J\!=\!3\!\rightarrow\!2$                      & \phantom{$<$}3.6(14)  & \phantom{$<$}5.5(14) &\phantom{$<$}5.9(14) & \phantom{$<$}1.2(15)\\
 HCO$^{+}$        & $J\!=\!4\!\rightarrow\!3$                      & \phantom{$<$}3.3(11)  & \phantom{$<$}4.4(12) & \phantom{$<$}9.4(11)& \phantom{$<$}1.0(12)\\
 HCO$^{+}$        & H$^{13}$CO$^{+}$ $J\!=\!4\!\rightarrow\!3$     & \multicolumn{1}{c}{...}  & \phantom{$<$}1.2(13) & \multicolumn{1}{c}{...}  & \multicolumn{1}{c}{...}  \\
 H$^{13}$CO$^{+}$ &  $J\!=\!4\!\rightarrow\!3$                     &           $<$1.5(11)  & \phantom{$<$}2.0(11) & \multicolumn{1}{c}{...} & \multicolumn{1}{c}{...}\\
 DCO$^{+}$        & $J\!=\!5\!\rightarrow\!4$                      & $<$2.9(11)  & \phantom{$<$}4.4(11) & \multicolumn{1}{c}{...} & \multicolumn{1}{c}{...}\\
 CN               & $J\!=\!3\frac{7}{2}\!\rightarrow\!2\frac{5}{2}$& \phantom{$<$}1.5(13)  & \phantom{$<$}6.6(13) &\phantom{$<$}1.5(13) & \phantom{$<$}1.5(13)\\
 HCN              & $J\!=\!4\!\rightarrow\!3$                      & \phantom{$<$}1.8(12)  & \phantom{$<$}9.2(12) & $<$1.0(12) & $<$1.2(12)\\
 H$^{13}$CN       & $J\!=\!4\!\rightarrow\!3$                      & \multicolumn{1}{c}{...}  & $<$4.8(11) &\multicolumn{1}{c}{...} &\multicolumn{1}{c}{...}\\
 HNC              & $J\!=\!4\!\rightarrow\!3$&             \multicolumn{1}{c}{...}   & $<$1.4(12)&\multicolumn{1}{c}{...}&\multicolumn{1}{c}{...}\\
 DCN              & $J\!=\!5\!\rightarrow\!4$&             \multicolumn{1}{c}{...}   & $<$4.0(10) &\multicolumn{1}{c}{...}&\multicolumn{1}{c}{...}\\
 CS               & $J\!=\!7\!\rightarrow\!6$&             \phantom{$<$}5.1(12)   & \multicolumn{1}{c}{...} &\multicolumn{1}{c}{...} &\multicolumn{1}{c}{...}\\
 \noalign{\smallskip}
 \hline
 \noalign{\smallskip}
 H$_{\rm 2}$CO    &$J\!=\!2_{12}\!\rightarrow\!1_{11}$ & \phantom{$<$}5.1(12) & \multicolumn{1}{c}{...}          &$<$2.1(12) &$<$2.7(13)\\
 H$_{\rm 2}$CO    &$J\!=\!3_{03}\!\rightarrow\!2_{02}$ & \phantom{$<$}1.7(12) & \multicolumn{1}{c}{...}          &$<$2.6(12) & \multicolumn{1}{c}{...}\\
 H$_{\rm 2}$CO    &$J\!=\!3_{22}\!\rightarrow\!2_{21}$ & $<$1.4(13)& \multicolumn{1}{c}{...}     &$<$9.4(12) & \multicolumn{1}{c}{...}\\
 H$_{\rm 2}$CO    &$J\!=\!3_{12}\!\rightarrow\!2_{11}$ & \phantom{$<$}7.1(11) & \multicolumn{1}{c}{...}          &$<$1.5(12) &$<$9.4(11) \\

 H$_{\rm 2}$CO    &$J\!=\!5_{15}\!\rightarrow\!4_{14}$ & \phantom{$<$}2.4(12) & $<$8.0(11)& $<$1.1(12) & $<$1.5(12)\\
 \noalign{\smallskip}
 \hline
 \noalign{\smallskip}
 CH$_{\rm 3}$OH   &$J\!=\!2_K\!\rightarrow\!1_K$ &   $<$7.1(13)         & \multicolumn{1}{c}{...}          & $<$2.8(13)&\multicolumn{1}{c}{...} \\
 CH$_{\rm 3}$OH   &$J\!=\!4_{\rm 2}\!\rightarrow\!3_1$ E$^+$ & $<$4.3(14)    &  \multicolumn{1}{c}{...}    & \multicolumn{1}{c}{...}
  &$<$6.9(13)\\   
 CH$_{\rm 3}$OH   &$J\!=\!5_K\!\rightarrow\!4_K$ &  $<$2.4(13)     &  \multicolumn{1}{c}{...}  & \multicolumn{1}{c}{...} &\multicolumn{1}{c}{...} \\
 CH$_{\rm 3}$OH   &$J\!=\!7_K\!\rightarrow\!6_K$ & \multicolumn{1}{c}{...} & $<$1.1(13)&\multicolumn{1}{c}{...} & \multicolumn{1}{c}{...}\\
 \noalign{\smallskip}
 \hline
 \noalign{\smallskip}
 N$_{\rm 2}$H$^+$          &$J\!=\!4\!\rightarrow\!3$ &             $<$1.4(12)   & $<$1.0(13) & \multicolumn{1}{c}{...} & $<$1.5(12) \\
 H$_{\rm 2}$D$^+$       &$J\!=\!1_{10}\!\rightarrow\!1_{11}$ &   $<$8.7(11)   & $<$4.4(12) & \multicolumn{1}{c}{...} & $<$1.0(12)\\
 \noalign{\smallskip}
 \hline
 \noalign{\smallskip}
 SO & $J\!=\!8_8\!\rightarrow\!7_7$ &\multicolumn{1}{c}{...}& $<$4.3(12)&\multicolumn{1}{c}{...}&\multicolumn{1}{c}{...} \\
\noalign{\smallskip}
 \hline
 \end{tabular}
 \end{center}
 \begin{flushleft}
   {\em Note.--}
   The excitation temperature is assumed to be 25~K for all lines;
  uncertainties in the column densities are of the order of 30\%, not
including uncertainties in the disk size. \\
   a(b) means $a \times$ 10$^b$. \\
   $^a$ $^{12}$CO column density derived from $^{13}$CO intensity assuming [$^{12}$C]/[$^{13}$C]=60. \\
 \end{flushleft}
 \end{table*}

 \begin{table*}
  \centering
 \caption{Beam-averaged molecular abundances with respect to H$_{\rm 2}$
 for the adopted disk sizes.\label{tab:mol_abundances}}
 \begin{tabular}{llllllll}
 \hline
\hline
  &  \multicolumn{7}{c}{$X=N/ N({\rm H_{\rm 2}})$}\\
 \noalign{\smallskip}
 \cline{2-8}
 \noalign{\smallskip}
 \multicolumn{1}{c}{Species} &
  \multicolumn{1}{c}{LkCa15}&\multicolumn{1}{c}{TW
  Hya}&\multicolumn{1}{c}{\object{HD~163296}}&\multicolumn{1}{c}{MWC
  480}&\multicolumn{2}{c}{DM~Tau}&\multicolumn{1}{c}{\object{IRAS 16293-2422}$^b$}\\ 
 \cline{6-7}
 & & & & & This work$^a$ & Dutrey et al. & \\
 \hline
 CO               &   \phantom{$<$}3.4(-07)     & \phantom{$<$}5.7(-08) & \phantom{$<$}3.1(-07)  & \phantom{$<$}6.9(-07) &  \phantom{$<$}9.6(-06)  & \phantom{$<$}1.5(-05) &\phantom{$<$}4.0(-05)\\
 HCO$^{+}$        &   \phantom{$<$}5.6(-12)     & \phantom{$<$}2.2(-11)$^c$  & \phantom{$<$}7.8(-12) & \phantom{$<$}1.0(-10) &\phantom{$<$}7.4(-10)&\phantom{$<$}7.4(-10)  &\phantom{$<$}1.4(-09)\\ 
 H$^{13}$CO$^{+}$ &    $<$2.6(-12) & \phantom{$<$}3.6(-13)  & \multicolumn{1}{c}{...}  &\multicolumn{1}{c}{...} &$<$3.6(-11) &$<$3.6(-11) &\phantom{$<$}2.4(-11)\\
 DCO$^{+}$        &    $<$2.31(-12)& \phantom{$<$}7.8(-13)  & \multicolumn{1}{c}{...} & \multicolumn{1}{c}{...}&\multicolumn{1}{c}{...} &\multicolumn{1}{c}{...} &\phantom{$<$}1.3(-11)\\ 
 CN               &   \phantom{$<$}2.4(-10)     & \phantom{$<$}1.2(-10)  & \phantom{$<$}1.3(-10)   &\phantom{$<$}1.4(-10) &\phantom{$<$}9.0(-09) &\phantom{$<$}3.2(-09) &\phantom{$<$}8.0(-11)\\ 
 HCN              &   \phantom{$<$}3.1(-11)     & \phantom{$<$}1.6(-11)  & $<$9.1(-12) & $<$1.1(-11) &\phantom{$<$}4.9(-10) &\phantom{$<$}5.5(-10) &\phantom{$<$}1.1(-09)\\ 
 H$^{13}$CN       &    \multicolumn{1}{c}{...}     & $<$8.4(-13)  & \multicolumn{1}{c}{...} &\multicolumn{1}{c}{...}   & \multicolumn{1}{c}{...} &\multicolumn{1}{c}{...} & \phantom{$<$}1.8(-11)\\
 HNC              &    \multicolumn{1}{c}{...}     & $<$2.6(-12) & \multicolumn{1}{c}{...}      & \multicolumn{1}{c}{...}&\phantom{$<$}1.5(-10) &\phantom{$<$}2.4(-10)&\phantom{$<$}6.9(-11)\\  
 DCN              &    \multicolumn{1}{c}{...}     &  $<$7.1(-14) & \multicolumn{1}{c}{...}      & \multicolumn{1}{c}{...}&\multicolumn{1}{c}{...} &\multicolumn{1}{c}{...} &\phantom{$<$}1.3(-11)\\ 
 CS               &   $<$8.5(-11)     &  \multicolumn{1}{c}{...}     & \multicolumn{1}{c}{...}      & \multicolumn{1}{c}{...}&\phantom{$<$}2.4(-10)&\phantom{$<$}3.3(-10)&\phantom{$<$}3.0(-09)\\ 
 H$_{\rm 2}$CO    &  \phantom{$<$}4.1(-11)      & $<$1.4(-12)  &  $<$1.0(-11) & $<$1.4(-11) &\phantom{$<$}2.4(-10)&\phantom{$<$}5.0(-10)& \phantom{$<$}7.0(-10)\\ 
 CH$_{\rm 3}$OH   &    $<$3.7(-10) & $<$1.9(-11)  &  $<$1.5(-10)  & $<$2.0(-09) &\multicolumn{1}{c}{...} &\multicolumn{1}{c}{...}          &\phantom{$<$}3.5(-10)\\ 
 N$_{\rm 2}$H$^+$ &    $<$2.3(-11) & $<$1.8(-11)  &  \multicolumn{1}{c}{...}     &$<$1.5(-11)&$<$5.0(-09) &$<$2.0(-10)& \multicolumn{1}{c}{...}\\ 
 H$_{\rm 2}$D$^+$ &    $<$1.5(-11) & $<$7.8(-12)  &  \multicolumn{1}{c}{...}     & $<$1.0(-11)& \multicolumn{1}{c}{...} &\multicolumn{1}{c}{...} & \multicolumn{1}{c}{...}\\ 
 SO               &\multicolumn{1}{c}{...} &$<$4.1(-11)&\multicolumn{1}{c}{...}&\multicolumn{1}{c}{...}&\multicolumn{1}{c}{...}&\multicolumn{1}{c}{...} & \phantom{$<$}4.4(-09)\\
 \hline
 \end{tabular}
\begin{flushleft}
   $^a$ Re-analysis of data from Dutrey et al.\ (\cite{Dutrey97}), see text\\ 
   $^b$ Outer envelope abundances from Sch\"oier et al.\ (\cite{Schoier02})\\
   $^c$ Value inferred from H$^{13}$CO$^+$ 
   \\
 \end{flushleft}
 \end{table*}
\begin{table*}
\centering
\caption{Relative molecular abundances \label{tab:abun_rel}}
\begin{tabular}{llll}
\hline
\hline
\noalign{\smallskip}
\multicolumn{1}{c}{Source} & \multicolumn{1}{c}{$N$(HCO$^+$)} & \multicolumn{1}{c}{$N$(CN)}&\multicolumn{1}{c}{$N$(DCO$^+$)}\\
\cline{2-2}
\cline{3-3}
\cline{4-4}
\noalign{\smallskip}
                            & \multicolumn{1}{c}{$N$(CO)}      & \multicolumn{1}{c}{$N$(HCN)}&\multicolumn{1}{c}{$N$(HCO$^+$)}\\
\hline
\object{LkCa~15}           & \phantom{$>$}1.6(-5) & \phantom{$>$}\ 7.9 & $<$0.411\\
\object{TW~Hya}            & \phantom{$>$}3.8(-4)$^1$ & \phantom{$>$}\ 7.1& \phantom{$>$}0.035\\
                           & \phantom{$>$}{\bf 1.4(-4)$^2$} &  &\\

DM~Tau$^a$        & \phantom{$>$}5.3(-5) & \phantom{$>$}\ 5.8&\\
DM~Tau$^b$        & \phantom{$>$}7.6(-5) & \phantom{$>$}18.4&\\
\hline
\object{HD~163296}         & \phantom{$>$}2.7(-5) & $>$12.4&\\
\object{MWC~480}           & \phantom{$>$}1.5(-5) & $>$11.7&\\
\hline
\object{IRAS 16293-2422}$^c$   & \phantom{$>$}3.6(-5) & \phantom{$>$}\ 0.07& \\
TMC-1$^d$             & \phantom{$>$}1.0(-4) & \phantom{$>$}\ 1.5&\\
Orion Bar$^e$         & \phantom{$>$}2.0(-5) & \phantom{$>$}\ 3.8&\\
IC~63$^f$             & \phantom{$>$}2.7(-5) & \phantom{$>$}\ 0.7&\\
\hline
\end{tabular}
\begin{flushleft}
  References.\ (a) Dutrey et al.\ (\cite{Dutrey97}); (b) from Table 7;
  (c) Sch\"oier et al.\ (2002); (d) Ohishi et al.\ (\cite{Ohishi92});
  (e) Hogerheijde et al.\ (\cite{Hogerheijde95}); (f) Jansen et al.\ 
  (\cite{Jansen95}). \\
{\em Notes.}\ $^1$ Value derived from
  H$^{13}$CO$^+$, assuming [$^{12}$C]/[$^{13}$C]=60.  
$^2$ Value derived from HCO$^+$.
\end{flushleft}
\end{table*}


 \section{Discussion}
 \label{moldisks:discussion}

\subsection{Molecular depletion}

Two processes have been put forward to explain the low molecular
abundances in disks.  First, in the disk midplane, the dust
temperature is so low ($<$~20~K) and the density so high ($n_{\rm
  H}>$~10$^9$ cm$^{-3}$) that most molecules including CO are frozen
onto the grain surfaces. This possibility is supported by the
detection of large amounts of solid CO at infrared wavelengths in the
disk around the younger class~I object \object{CRBR 2422.8--3423} (Thi
et al.\ \cite{Thi02b}).  In this environment, surface chemistry can
occur but the newly-formed species stay in the solid phase and thus
remain unobservable at millimeter wavelengths, except for a small
fraction which may be removed back in the gas phase by non-thermal
desorption processes such as cosmic-ray spot heating.

 Second, the photodissociation of molecules in the upper layers of
 protoplanetary disks by the ultraviolet radiation from the central
 star and from the ambient interstellar medium can limit the lifetime
 of molecules. The ultraviolet flux from the central star can reach
 10$^4$ times the interstellar flux (Glassgold, Feigelson \& Montmerle
 \cite{Glassgold00}). Aikawa et al.\ (\cite{Aikawa2002}) and van
 Zadelhoff et al.\ (\cite{Zadelhoff03}) have modeled the chemistry in
 disks, taking these mechanisms into account.  Their models show that
 molecules are abundant in the intermediate height regions of disks,
 consistent with the derived temperature range (20--40~K) for the
 emitting gas. According to the flaring disk model, this intermediate
 region is located just below the warm upper layer
 ($T\simeq$~100~K). 

 The molecular abundance distributions predicted by the above chemical
 models including photodissociation and freeze-out have been put into
 a 2-D radiative transfer code to compute the level populations using
 statistical equilibrium rather than LTE and to take the optical depth
 effects properly into account. The resulting integrated fluxes can be
 compared directly with observations. As shown by Aikawa et al.\
 (\cite{Aikawa2002}) they differ by factors of a few up to an order of
 magnitude, which indicates that such models are to first order
 consistent with the data.


\subsection{CN/HCN abundance ratio}

Table~\ref{tab:abun_rel} includes the CN/HCN abundance ratios derived
for the disks.  Compared with \object{IRAS 16293-2422}, the CN/HCN
ratio is more than two orders of magnitude larger, and even compared
with galactic PDRs, all disk ratios are higher.  The disk ratios may
be overestimated due to underestimate of HCN optical depth effects,
but the high values are a strong indication that photodissociation
processes play a role in the upper layers of the disks. 

CN is particularly enhanced by photochemistry since it is produced by
radical reactions involving atomic C and N in the upper layers as well
as photodissociation of HCN.  Moreover, CN cannot be easily
photodissociated itself since very high energy photons ($<$1000 \AA,
$>$ 12.4 eV) are required to destroy the radical (van Dishoeck
\cite{vD87}).  The CN/HCN ratio appears to be higher in disks around
Herbig~Ae stars than around T~Tauri stars, although our high ratio of
$> 11$ for MWC 480 disagrees with the ratio of $\sim$4 by Qi
(\cite{Qi01}) with OVRO.  The disagreement between the results of Qi
(\cite{Qi01}) and ours can be ascribed to the fact that Qi (2001) had
H$^{13}$CN data available to constrain the optical depth of the HCN
line.  Also, their 1--0 lines are less sensitive to the adopted disk
density structure.  In general the fluxes from MWC 480 for transitions
which have high critical densities are lower than those for T~Tauri
stars, whereas the CO fluxes (with lower critical densities) are
higher. This may imply that the level populations are subthermal for
the disks around Herbig Ae stars.

Van Zadelhoff et al.\ (\cite{Zadelhoff03}) have investigated the
effects of different UV radiation fields on the disk chemistry,
focusing on T Tauri stars with and without excess UV emission.  CN is
clearly enhanced in the upper disk layers for radiation fields without
any excess UV emission owing to its reduced photodissociation.  When
convolved with the JCMT beams, however, the differences are difficult
to discern: the HCN emission is nearly identical for the different
radiation fields, whereas the CN emission varies by only a factor of a
few. Since T Tauri stars like TW~Hya have been observed to have excess
UV emission (Costa et al.\ \cite{Costa2000}) ---probably originating
from a hot boundary layer between the accretion disk and the star---,
the difference in the CN/HCN chemistry with the Herbig Ae stars may be
smaller than thought on the basis of just the stellar spectra. Bergin
et al.\ (\cite{Bergin03}) suggest that strong Ly$\alpha$ emission
dominates the photodissociation rather than an enhanced continuum
flux. Since CN cannot be photodissociated by Ly$\alpha$ radiation but
HCN can (Bergin et al.\ 2003, van Zadelhoff et al.\ 2003), the CN/HCN
ratio is naturally enhanced.

Other chemical factors can also affect the CN/HCN ratio. Radicals such
as CN are mainly destroyed by atomic oxygen in the gas-phase and
therefore a lower oxygen abundance can increase the CN/HCN ratio.
Since atomic oxygen is a major coolant for the gas, a lower abundance
will also maintain a higher mean kinetic temperature.
Alternatively, the dust temperature could be in the regime that
HCN is frozen out but CN not because the two
molecules have very different desorption energies ($E_{\rm
  des}$(CN)=1510~K and $E_{\rm des}$(HCN)=4170~K; Aikawa et
al.~\cite{Aikawa97}).

Yet an alternative explanation for high CN abundances is production by
X-ray photons emitted from the active atmosphere of T~Tauri stars
(e.g., Aikawa \& Herbst \cite{Aikawa99a}; Lepp \& Dalgarno
\cite{Lepp96}). \object{TW~Hya} is a particularly strong X-ray
emitter, with a measured X-ray flux 10 times higher than the mean
X-ray flux observed toward other T~Tauri stars (Kastner et al.\ 
\cite{Kastner02}). In addition, H$_{\rm 2}$ $v\!=\!1\!\rightarrow\!0$
S(1), another diagnostic line of energetic events, has been observed
toward this object (Weintraub, Kastner, \& Bary\ \cite{Weintraub00}).
\object{TW~Hya} may however constitute a special case since neither
LkCa~15 nor DM~Tau seems to show enhanced X-ray emission, yet they
have a similar CN/HCN ratio. Further observations of molecules in
disks around strong X-ray emitting pre-main-sequence stars are
warranted to better constrain the contribution of X-rays on the
chemistry in disks.

\subsection{HCO$^+$/CO abundance ratio}

Table~\ref{tab:abun_rel} compares the HCO$^+$/CO ratios found in the
disks with those found in a protostellar region (\object{IRAS
  16293-2422}), a dark cloud (\object{TMC-1}) and two galactic
photon-dominated regions (PDRs) (\object{Orion Bar} and
\object{IC~63}). Within a factor of two, all values are very similar,
except for the \object{TW~Hya} disk.  It should be noted, however,
that except for TW~Hya, the ratios in disks have been derived from the
optically thick HCO$^+$ line and may therefore be underestimates.
Indeed, the ratio obtained using the main HCO$^+$ isotope for
\object{TW~Hya} is closer to that of the other objects. Observations
of H$^{13}$CO$^+$ for all disks are warranted to make definitive
conclusions.

HCO$^+$ is produced mainly by the gas phase reaction H$_3^+$ + CO
$\rightarrow$ HCO$^+$ + H$_2$.  Its formation is increased by enhanced
ionization (e.g., by X-ray ionization to form H$_3^+$ in addition to
cosmic rays) and by enhanced depletion (which also enhances H$_3^+$,
see e.g., Rawlings et al.\ \cite{Rawlings92}). The fact that all
HCO$^+$ abundances in disks are higher than those found in normal
clouds (after correction for HCO$^+$ optical depths) suggests that
these processes may play a large role in the intermediate warm disk
layer where both molecules are thought to exist. In this context it is
interesting to note that TW~Hya has the largest depletion of CO and is
also the most active X-ray emitter (see below). 

The derived HCO$^+$ abundances provide a lower limit to the ionization
fraction in disks. The typical values of $10^{-11}-10^{-10}$ are high
enough for the magnetorotational instability to occur and thus provide
a source of turbulence and mixing in the disk (e.g., Nomura
\cite{Nomura2002}).  Ceccarelli et al. ~(\cite{Ceccarelli04}) used
their H$_2$D$^{+}$ observations toward \object{DM~Tau} and {TW~Hya} to
derive an electron abundance of (4--7). 10$^{-10}$, assuming that
H$_3^+$ and H$_2$D$^{+}$ are the most abundant ions.  These values
refer to the midplane of those disks where the depletion of CO
enhances the H$_2$D$^{+}$/H$_3^+$ ratio (see \S~5.5), whereas our
values apply to the intermediate layer where most of the HCO$^+$
emission arises.

\subsection{H$_{\textsf 2}$CO/CH$_{\textsf 3}$OH}

Lines of H$_{\rm 2}$CO have been detected toward only one source,
\object{LkCa~15}. CH$_{\rm 3}$OH is not detected in the single-dish
observations of any disks, although it is seen in the OVRO
interferometer data toward \object{LkCa~15} by Qi (2001), who derived
a CH$_{\rm 3}$OH column density of 7--20 $\times$ 10$^{14}$ cm$^{-2}$
compared with our upper limit of 9.4 $\times$ 10$^{14}$ cm$^{-2}$.
Our upper limits for H$_{\rm 2}$CO and CH$_3$OH are derived from the
spectra with the lowest r.m.s. Methanol has been detected from the
class 0 protostar L1157 by Goldsmith, Langer \& Velusamy
(\cite{Goldsmith99}), where its emission has been ascribed to a
circumstellar disk. However, this object is much younger than those
studied here and presumably has a different physical structure and
chemical history.

The H$_{\rm 2}$CO/CH$_{\rm 3}$OH abundance ratio of
$>$~0.15 for \object{LkCa~15} is consistent with values found for
embedded YSOs (see Table 8 of Sch\"oier et al.  \cite{Schoier02} for
IRAS 16293--2422 and van der Tak et al.\ \cite{vdT00} for the case of
massive protostars).  Heating of the disk, whether by ultraviolet- or
X-rays, should lead to strong ice evaporation and thus to enhanced
gas-phase abundances for grain-surface products. Both CH$_{\rm 3}$OH
and H$_2$CO have been detected in icy mantles, with the CH$_3$OH
abundance varying strongly from source to source (Dartois et al.\ 
1999, Keane et al.\ \cite{Keane01}, Pontoppidan et al.
\cite{Pontop03}). For the few sources for which both species have been
seen, the solid H$_2$CO/CH$_3$OH ratio varies from 0.1--1.  Thus, the
observed ratio in LkCa~15 could be consistent with grain surface
formation of both species.  Since their absolute abundances are much
lower than typical ice mantle abundances of $10^{-6}$ with respect to
H$_2$, this indicates that most of the CH$_{\rm 3}$OH and H$_2$CO, if
present, is frozen onto grains.  Deeper searches for both species in
disks are warranted.

Protoplanetary disks are places where comets may form and their
volatile composition may provide constraints on their formation
models.  The CH$_3$OH abundance is known to vary significantly between
comets.  For example, comet C/1999 H1 (Lee) shows a CO/CH$_{\rm 3}$OH
ratio around 1 whereas Hale-Bopp and Hyakutake have ratios of 10 and
14 respectively (Biver et al.\ \cite{Biver00}). Comet Lee probably
belongs to the so-called $``$methanol-rich comets$"$ group
(Bockel\'{e}e-Morvan, Brooke \& Crovisier \cite{Bockelee98}; Davies et
al.\ \cite{Davies93}).  In addition, the measured CO abundance is
$\sim$1.8$\pm$ 0.2\% compared to H$_{\rm 2}$O, 5 times less than found
in Hale-Bopp.  Alternatively, Mumma et al.\ (\cite{Mumma01}) propose
that Comet Lee has been heated sufficiently after its formation for CO
to evaporate but not CH$_{\rm 3}$OH, so that CH$_{\rm 3}$OH abundance
is not enhanced but rather CO is depleted. Mumma et al.\
(\cite{Mumma01}) notice that the CH$_{\rm 3}$OH/H$_{\rm 2}$O and
CO/H$_{\rm 2}$O ratios vary strongly among comets coming from the
giant-planets regions. The picture is not complete since CO can be
converted to CO$_{\rm 2}$, whose abundance is high in interstellar
ices (e.g., Ehrenfreund \& Charnley \cite{Ehrenfreund00}) but less
well known in comets (10\% in comet 22P/Kopff, Crovisier et al.\
\cite{Crovisier99}).

Long period comets were probably formed in the Jupiter-Saturn region
(around 5--20~AU), whereas our data are only sensitive to distances of
more than 50~AU.  It would therefore be more relevant to compare the
composition of protoplanetary disks to that of Kuiper Belt Objects,
which were formed beyond 50~AU in the solar nebula. The chemical
composition of Kuiper Belt Objects is not well known (see Jewitt \&
Luu \cite{Jewitt00}), although observations show that comet nuclei and
Kuiper Belt Objects have different surface compositions (Luu \& Jewitt
\cite{LuuJewiit02}; Jewitt \cite{Jewitt02}). The nature of Centaur
objects is better understood. It is believed that Centaur objects were
formed beyond 50~AU and recently entered the planetary zone with
orbits crossing those of the outer planets. The best studied Centaur
object, 5145 Pholus, shows the presence of CH$_{\rm 3}$OH although the
exact amount is not well constrained (Cruikshank et al.\ 
\cite{Cruikshank98}).


\subsection{D/H ratio in a circumstellar disk}

The D/H ratio in the \object{TW~Hya} disk of 0.035 $\pm$0.01 has been
derived from the H$^{13}$CO$^+$ and DCO$^+$ column density ratios,
assuming an isotopic ratio [$^{12}$C]/[$^{13}$C] of 60 (van Dishoeck
et al.\ \cite{Dishoeck2003}).  Hints of H$^{13}$CN and DCN features
are seen in the \object{TW~Hya} spectra, but neither of them is
definitely detected. Searches for other deuterated species in the
LkCa~15 disk, in particular DCN and HDO, are reported by Kessler et
al.\ (\cite{Kessler03}).
  
It should be kept in mind that our observations provide only a value
of the D/H ratio averaged over the entire disk. Models of the
DCO$^+$/HCO$^+$ abundance ratio show that it remains rather constant
throughout the disk down to a radius of 100 AU, however (Aikawa et
al.\ \cite{Aikawa2002}).  All values are significantly higher than the
elemental [D]/[H] abundance of 1.5$\times 10^{-5}$ (Pettini \& Bowen
\cite{Pettini01}; Moos et al.\ \cite{Moos02}).
  
  Theoretically, the amount of deuterium fractionation in molecules
  depends on the gas kinetic temperature, which drives the isotopic
  exchange reactions, and on the cosmic ray ionization rate (Aikawa \&
  Herbst \cite{Aikawa99b}).  Also, the abundance is enhanced if CO is
  significantly depleted onto grains (Brown \& Millar \cite{Brown89}).
  Thus, the amount of deuterium fractionation can serve as a tracer of
  the temperature history of the gas.  The deuterium fractionation can
  be further enhanced by grain-surface formation (Tielens
  \cite{Tielens83}), although not for DCO$^+$/HCO$^+$. Recent chemical
  models succeed in explaining the high fractionation observed here
  and in dark cloud cores (Rodgers \& Millar \cite{Rodgers96}; Roberts
  \& Millar \cite{Roberts00}, Tin\'{e} et al.\ \cite{Tine00}), but
  only if significant freeze-out is included (Roberts et al.\ 
  \cite{Roberts02}, \cite{Roberts03}).  
Our observed values are also close to those found in disk models which
include a realistic 2D temperature and density profile with freeze-out
(Aikawa et al.\ \cite{Aikawa2002}).

Table~2 in van Dishoeck et al.\ (2003) compares the D/H ratio found in
disks to typical values for the D/H ratio in different protostellar
and cometary environments.  The value found in disks is somewhat
higher than that in the low-mass protostellar envelope of IRAS
16293--2422, but comparable to that seen in dark cloud cores.
DCO$^+$/HCO$^+$ has not been observed in comets, but the D/H ratios
derived from DCN/HCN in pristine material in jets originating from
below the comet surface is found to be similar to that seen for
DCO$^+$/HCO$^+$ in the TW~Hya disk (Blake et al.\ \cite{Blake99}).
Alternatively, Rodgers \& Charnley (\cite{Rodgers02}) propose that the
DCN and HCN seen in these cometary jets are the photodestruction
products of large organic molecules or dust grains. In either case,
the D/H ratio of pristine icy material in comets is high.  The
similarity suggests that either the gas is kept at low temperatures as
it is transported from pre-stellar clouds to disks and eventually to
comet-forming regions, or, alternatively, that the DCO$^+$/HCO$^+$
ratio has been reset by the chemistry in disks. Comparison with D/H
ratios of molecules which likely enter the disks as ices are needed to
distinguish these scenarios.


\section{Conclusions}

We surveyed low- and high-$J$ transitions of simple organic molecules
in two classical T~Tauri and two Herbig~Ae stars. Analysis of line
ratios indicates that the emission comes from a dense ($n_{\rm H}$ =
10$^6$--10$^8$ cm$^{-3}$) and moderately warm region ($T\simeq$
20--40~K).  Detailed fits to the $^{12}$CO 3--2 emission line profiles
provide independent estimates of the sizes of the disks.

Emission from the ion HCO$^+$ and the radical CN are particularly
strong, indicating an active gas-phase chemistry in the surface layers
of disks which is affected by UV radiation from the central stars.
H$_{\rm 2}$CO is detected in one source but CH$_{\rm 3}$OH is not
observed in any object in our sample.  In one source (\object{TW~Hya})
the detection of DCO$^+$ allows to constrain the DCO$^+$/HCO$^+$ ratio
to $\sim$0.035, a value that is higher than that found in the
envelopes of low-mass protostars but comparable to that observed in
cold dark cores, where fractionation due to low temperature chemistry
and CO freeze-out is important.
 
This work demonstrates that organic chemistry in disks around low- and
intermediate-mass pre-main-sequence stars can now be studied
observationally.  The detection of molecular species in disks is
hampered, however, by the small sizes of disks compared with the
actual beams of single-dish telescopes. Moreover, because the total
amount of material is small (few $\times$ 10$^{-2}$ M$_{\sun}$), the
observations are limited to the most abundant species.  It is likely
that the chemistry in more tenuous disks, in which the ultraviolet
radiation can penetrate through the entire disk, is different from
that for our objects (e.g., Kamp \& Bertoldi
\cite{Kamp00}, Kamp et al.\ \cite{Kamp03}). 
Although the outer disks can be resolved by current
millimeter interferometers, integration times are too long to do
molecular line surveys and the inner tens of AU are still out of
reach.  The detection of more complex and much less abundant molecules
in protoplanetary disks at different stages of evolution awaits the
availability of the {\it Atacama Large Millimeter Array}
({ALMA}). Complementary infrared observations of solid-state species
along the line of sight of edge-on protoplanetary disks will help to
constrain quantitatively the level of depletion in the mid-plane of
disks.

 \begin{acknowledgements} \label{moldisks:acknowledgements} WFT thanks
   PPARC for a Postdoctoral grant to UCL.  This work was supported by
   a Spinoza grant from the Netherlands Organization for Scientific
   Research to EvD and a postdoctoral grant (614.041.005) to WFT.  We
   thank Remo Tilanus, Fred Baas, Michiel Hogerheijde, Kirsten
   Knudsen-Kraiberg, Annemieke Boonman, and Peter Papadopoulos, who
   have performed some of the JCMT observations in service; Geoff
   Blake, Charlie Qi and Jackie Kessler for communicating their OVRO
   results prior to publication; and Yuri Aikawa for fruitful
   discussions on disk models. We acknowledge the IRAM staff at
   Granada for carrying part of the observations in service mode.
\end{acknowledgements}


\end{document}